%% file: SwitchPairing2019.tex
\documentclass[10pt,journal,compsoc]{IEEEtran}

\usepackage{booktabs} 

\usepackage{amssymb}
\usepackage{amsmath}
\usepackage{diagbox}
\usepackage{color}
\usepackage[utf8x]{inputenc}
\usepackage{indentfirst}
\usepackage{hyperref}
 
 \usepackage{graphicx}
 \usepackage{graphicx}
\usepackage{color}
\usepackage[justification=centering]{caption}
\usepackage{caption}
 \usepackage[ruled]{algorithm2e}
\usepackage{multicol}
\usepackage{multirow}
\usepackage{subcaption}
\newcommand{\delete}[1]{}

\definecolor{mygray}{gray}{.9}
\usepackage{colortbl}
\ifCLASSOPTIONcompsoc
  \usepackage[nocompress]{cite}
\else
  \usepackage{cite}
\fi

 \newcommand{\nop}[1]{}
\newcolumntype{P}[1]{>{\centering\arraybackslash}p{#1}}
\begin{document}

\title{Peripheral-free Device Pairing by Randomly Switching Power }

 \author{Zhijian Shao, 
        Jian Weng, ~\IEEEmembership{Member,~IEEE}
       Yue Zhang,
       Yongdong Wu,
       Ming Li,\\
       Jiasi Weng,
        Weiqi Luo,
        Shui Yu ~\IEEEmembership{Senior Member,~IEEE} 
\IEEEcompsocitemizethanks{\IEEEcompsocthanksitem
Zhijian Shao, Jian Weng,Yue Zhang, YongDong Wu, Jiasi Weng, Ming Li and Weiqi Luo are with the College of Informatin Science and Technology / College of Cyber security, National Joint Engineering Research Center of Network Security Detection and Protection Technology, and Guangdong Key Laboratory of Data Security and Privacy Preserving, Jinan University, Guangzhou, 510632, China; 

Shui Yu with the University of Technology Sydney, Ultimo, NSW, Australia

\IEEEcompsocthanksitem Jian Weng is the corresponding author : cryptjweng@gmail.com  ;}
 
\thanks{Manuscript received February, 2020; revised February, 2020.}}

\markboth{Journal of \LaTeX\ Class Files,~Vol.~14, No.~8, 2020}%
{Shell \MakeLowercase{\textit{et al.}}: Bare Demo of IEEEtran.cls for Computer Society Journals}
 
\IEEEtitleabstractindextext{%
\begin{abstract}

The popularity of Internet-of-Things (IoT) comes with security concerns. Attacks against wireless communication venues of IoT (e.g., Man-in-the-Middle attacks) have grown at an alarming rate over the past decade. Pairing, which allows the establishment of the secure communicating channels for IoT devices without a prior relationship, is thus a paramount capability. Existing secure pairing protocols require auxiliary equipment/peripheral (e.g., displays, speakers and sensors) to achieve authentication, which is unacceptable for low-priced devices such as smart lamps. 
This paper studies how to design a peripheral-free  secure pairing protocol. Concretely, 
we design the protocol,  termed SwitchPairing, via out-of-box power supplying chargers and on-board clocks,  achieving security and economics at the same time.
When a user wants to pair two or more devices, he/she connects the pairing devices to the same power source, and presses/releases the switch on/off button several times. Then, the press and release timing can be used to derive symmetric keys. We implement a prototype via two CC2640R2F development boards from Texas Instruments (TI) due to its prevalence. Extensive experiments and user studies are also conducted to benchmark our protocol in terms of efficiency and security.



\end{abstract}

}
 
\maketitle

\IEEEdisplaynontitleabstractindextext
 
\IEEEpeerreviewmaketitle

\input{Sections/sec1-introduction.tex}
 \input{Sections/sec2-relatedwork.tex}

\input{Sections/sec3-pairing.tex}
\input{Sections/sec4-switchpairing.tex}
\input{Sections/sec6-implemetation.tex}

\input{Sections/sec5-secanalysis.tex}
\input{Sections/sec7-evaluation.tex}
 \input{Sections/sec8-discussion.tex}

\input{Sections/sec9-conclusion.tex}

\input{Sections/sec-acknowledgement.tex}

\ifCLASSOPTIONcompsoc

\bibliographystyle{IEEEtran}
\bibliography{SwitchPairing2019}

\input{Sections/sec-bio}

\ifCLASSOPTIONcaptionsoff
  \newpage
\fi

\end{document}

%% file: Sections/sec1-introduction.tex
\section{Introduction}
Internet of Things (IoT) is a novel paradigm that has  received a lot of recent attention. It features a broad spectrum of applications, ranging from smart home appliances to smart city and vehicle to vehicle networks.  In the context of IoT, all things will be interconnected, and a user may control the things around via a smart controller such as a smartphone. IoT arguably will change the way we conduct our everyday affairs. Recent reports and statistics indicate that such an IoT era is approaching: according to a report from statistics \cite{IoTmarket}, more than 30 billion devices/things will be connected to the internet by the end of 2020.  This number will hit 75 billion by 2025.

However, there are various attacks that may threat the security of IoT, such as Man-in-the-Middle (MITM) attacks \cite{barcena2015insecurity} and spoofing attacks \cite{nawir2016internet}. Countermeasures have been proposed to hinder these attacks, and using pairing protocols is one of these countermeasures. An IoT system usually involves a cloud server, some IoT devices, and a remote controller such as a smartphone with specific apps. The communications between the cloud server and the remote controller/IoT devices may be secured by public-key cryptographic protocols such as Transport Layer Security (TLS) \cite{rescorla2006datagram}, while pairing is designed to build upon a secure channel for the IoT devices and the remote controller. One example of pairing protocol is Bluetooth Simple Secure Protocol (SSP) \cite{Zhang2019}, which enables two devices equipped with Bluetooth modules such as a remote controller and an IoT device in close proximity to communicate securely. 
 
 Pairing protocols that require the pairing devices in close proximity such as  Bluetooth SSP are referred to proximity-based device pairing protocol \cite{wu2017attack}. In these schemes, the owner is close to his pairing devices (i.e., the remote controller and the IoT devices) and may need to perform specific operations, while the attackers are further away that do not have physical access of them. To achieve the pairing protocols, trusted auxiliary channels such as audio-visual signal\cite{prasad2008efficient,saxena2006secure}, correlated human behavior\cite{chagnaadorj2013mimicgesture,4.2} will be involved to exchange secrets, which latter will be used to derive a security key on each pairing device. 
For example, in {\em Passkey-entry} pairing strategy of Bluetooth Low Energy (BLE) \cite{4.2}, one device (either the remote controller or the IoT device) displays a 6 digits pin (i.e. secrets), while the pairing operator sees the pin, and enters the pin to another device. 
In the {\em Numeric Comparison} pairing strategy of BLE, the owner has to check if pins displayed on the two pairing devices are the same and makes confirmation by pressing the ''YES'' buttons on both devices.

These proximity-based device pairing protocols are not omnipotent since it only gives a trade-off between security and economics. That is, manufacturers have to pay extra costs to guarantee the security of their products. For example, {\em Passkey-entry} or {\em Numeric Comparison} of BLE requires one device or the two devices to equip displays or keypads, while other protocols may require both of the two devices integrate out of band communication modules, such as NFC modules, speakers or cameras \cite{prasad2008efficient,saxena2006secure,miettinen2014context,han2017convoy}. Instead of spending extra cost to apply secure pairing strategies, the manufacturers may rather choose insecure pairing protocols such as {\em Just Works} of Bluetooth. For example, it is unacceptable for some low-priced devices, such as smart lamp to equip a display or keyboard only for pairing purpose. The fact that using the insecure protocols rather than the secure pairing protocols has grave impacts given IoT broad application domains, in which the products are plagued by MITM attacks.

This paper studies how to design a secure IoT pairing protocol that requires no extra cost for pairing devices, allowing devices to achieve security and economics simultaneously.
The term ``without extra cost'' in our context refers to a scenario where the pairing devices set up a trusted channel without auxiliary equipment such as displays or keypads.  
Our insight is that the owners of two devices need to assist the two pairing devices to exchange secrets securely. Therefore, auxiliary equipment is involved, such as a display or a keypad.  Instead of using auxiliary equipment, if these devices can establish alternative trusted channels by using their out-of-box components, like the power supply chargers, the extra cost can be avoided.
 
Motivated by our observations, we carefully investigate the procedure of secure pairing protocol and propose a new pairing protocol, termed  \texttt{SwitchPairing}. We achieve secure pairing via the power supply charger and on-board clock of the devices since almost every IoT device has these components. 
Our pairing protocol works as follows:
\begin{enumerate}
\item The user connects the pairing devices to the same power source, i.e., a plug;  
\item Two paring devices communicate with each other via the communicating channel such as Bluetooth channel and synchronize the frequency of their on-board clocks;
\item The user switches the power source on and off intentionally several times. The timing of switch on/off are recorded by each pairing devices, respectively;
\item The timing of the switch on/off are regarded as the initial values that can be used to derive the cryptographic keys and secure the communicating channel.
\end{enumerate}

The benefits of adopting our protocol are threefold: 
(i) The usage of the power supply charger and on-board clock will keep our design economically viable and suitable for all kinds of devices, even the ones as simple as smart lamps.
(ii) Our protocol is capable of thwarting attacks such as MITM attack and more secure when compared with other pairing protocols. In our pairing protocol, the attacker will fail to obtain the timing of power switch off/on, even if the attacker can peep the user's operations somehow. Contrastingly, the classic pairing protocol such as Bluetooth pairing protocols will become completely defenseless when such an assumption is made. Moreover, unlike the most of context-aware pairing protocol which will fail to maintain their security when an attacker is nearby, our protocol can defend the attackers without physical access to the device and the threat model is much more realistic. Finally, in terms of defending guessing attack, the exchanged secrets can be very strong, depending on the precision of the on-board clock and the times that the user switches on/off the plug (See Section \ref{subsec:analysis}). 
(iii) Our \texttt{SwitchPairing} can also be performed on multiple devices rather than just two devices, seeing that multiple devices can be connected to the same power source. In this case, all devices can follow the same procedure of our protocol without any modification, obtaining the same cryptographic key.


\textbf{Contributions}: We summarize the main contributions of this paper as follows:

\begin{itemize}

\item We design a new pairing protocol, \texttt{SwitchPairing}, in which the pairing devices can negotiate the same security key without any extra equipment. 
 Our design provides a good guideline for the community to follow in terms of designing economically viable and secure pairing protocols.   
\item We implement a prototype via two CC2640R2F developing boards from Texas Instruments (TI) due to the prevalence of their products and the support of the IoT programming framework. Implementation criteria is given, so that the community can benefit from our design in a timely manner. 
\item Extensive evaluations and security analysis have been performed to benchmark our protocol. We conclude that our new protocol can also achieve high security with low overloads. In addition, for those devices with abundant I/O capabilities, such as speakers, cameras, and displays, can also adopt our protocol to achieve their security requirements. 
 
\end{itemize}


\textbf{Roadmap}:   The rest of this paper is organized as follows. Related works are presented in Section \ref{sec:relatedwork}. Section \ref{sec:ble} provides the necessary background information of Pairing. In Section \ref{sec:design}, we illustrate our threat model and the design of our pairing protocol. The implementation criteria is elaborated in Section \ref{sec:implementation}. Section \ref{subsec:analysis} presents the security analysis of our Switchpairing protocol.  Evaluations and user studies are presented in Section \ref{sec:eva}. Section \ref{sec:disscuss} discusses a few interesting topics.
 This paper is concluded in Section \ref{sec:conclusion}.


%% file: Sections/sec2-relatedwork.tex
\section{Related Work} 
\label{sec:relatedwork}

In this Section, we first review secure pairing protocols, which evidences the added value that our paper brings to state of the art. We also discuss the attacks on pairing protocols. It is noteworthy that our protocol can also defense these attacks. 

\subsection{Secure pairing protocols}
\label{subsec:reviewpairing}

We first review the secure pairing protocols. Secure pairing can establish an association (e.g., generating the same key) for two or more devices without a prior relationship. All solutions below require trusted certification authorities, physical connection \cite{stajano1999resurrecting} (which may subject to specification of interface ) or auxiliary channels such as speakers \cite{soriente2008hapadep},  displays \cite{mccune2005seeing} and cameras \cite{kumar2009caveat}.   
 
\textbf{Trusted certification authorities}: In early times, secure pairing protocols are achieved by using the public-key based key exchange protocols such as authenticated Diffie-Hellman protocols \cite{krawczyk2003sigma}. In these schemes \cite{hoepman2005ephemeral,hoepman2004ephemeral}, the pairing devices must pre-share the same root CA as a trusted root. However, it is unrealistic for heterogeneous IoT devices to have the same root CA.  

\textbf{Visual/Audio-based pairing protocols}: In the scheme proposed by Mccune \emph{et al.} \cite{mccune2005seeing}, a device is required to display a barcode and users have to take a snapshot then share to the peer somehow. Their scheme requires devices to have cameras and displays.  Saxena \emph{et al.} \cite{saxena2006secure} improved this scheme. They demonstrated that mutual authentication can be achieved with an unidirectional visual channel. In their paper, one device encodes the secrets as light signals by turning the light source on and off repeatedly, while the other device uses its camera to receive these signals and decode the secrets. Their solution still requires one device to equip a camera, which is clearly a burden for devices like smart lamps. Additionally, there are various options for two pairing devices equipping displays to achieve secure pairing, such as hash value, strings, colorful flag \cite{kumar2009caveat}, and random figures \cite{perrig1999hash}, which requires users to hold on the two pairing devices and compare the characters or figures displayed. 
Particularly, Bluetooth pairing protocols (Passkey Entry and Numeric Comparison) are belongs to visual based pairing protocols. There are also pairing protocols established on audio channels, such as HAPADEP \cite{soriente2008hapadep} proposed by Soriente \emph{et al.} and work by Prasad \cite{prasad2008efficient}. In these protocols,  auxiliary equipment such as speakers is required on the pairing devices. 

 \textbf{Sensor-based pairing protocols}: With the development of IoT technologies, more advanced auxiliary channels have be involved in the pairing process, such as Near Field Communication (NFC) \cite{4.2}, Laser \cite{mayrhofer2007human},  Infrared \cite{balfanz2004network,kindberg2003secure}, accelerometers \cite{lester2004you}. Holmquist \emph{et al.} \cite{holmquist2001smart} proposed a pairing protocol that allows devices with vibration sensors to achieve the pairing securely. In their scheme, a user is required to take the two pairing devices in one hand and shakes them, so that the similar movement data will be sensed by the vibration sensors. Similar movement data can then be used to generate the same key. A similar solution was discussed in efforts by Mayrhofer \cite{mayrhofer2009shake}. Nevertheless,  low priced devices, such as lamps usually do not have vibration sensors.  
 More recently, secure pairing is made possible due to the popularity of various on-board sensors, such as magnetometers \cite{jin2015magpairing}, motion
detectors \cite{li2010group} and footstep monitors \cite{pan2015indoor}. Sometimes, a pairing protocol may involve multiple sensors, such as the work by Han \cite{han2018you}. Using sensors to achieve secure pairing without human involvement is referred to context-based pairing in pairing terminology. Chaotic signals \cite{haroun2015secret},  ambient noise and luminosity \cite{miettinen2014context,han2017convoy} can be used as secrets to generate symmetric keys.  It can be concluded that all these pairing protocols require auxiliary equipment.

\subsection{Attacks against pairing protocols}
\label{insecpair}

In this Section, we will review the efforts related to the attacks on pairing protocols.
Rudimentary pairing protocols are subject to various attacks, such as eavesdropping attack or PIN cracking \cite{becker2007bluetooth,spill2007bluesniff,munro2008breaking}. In 2008, Haataja \emph{et al.} \cite{haataja2008practical,haataja2008man} propose  MITM attacks against SSP of Bluetooth Classic. Their insights are a malicious device can manipulate their I/O capabilities, and pair with a victim device using Just Works method.  Early in 2012,  Gomez \emph{et al.} \cite{gomez2012overview} present that BLE suffers from some security issues, such as replay attacks.  Mike Ryan \emph{et al.} \cite{ryan2013bluetooth} show  that for LE Legacy which is the early version of BLE, the Passkey Entry is vulnerable to guessing attacks. To explore the vulnerabilities of LE Legacy pairing, he also releases a tool named crackle. Based on this principle, Rosa \emph{et al.} \cite{rosa2013bypassing} and Zegeye \emph{et al.} \cite{zegeye2015exploiting} design their own attack vectors that break the LE Legacy pairing effectively. Da-zhi \emph{et al.} \cite{dazhisun} demonstrate that reusing a Passkey in Passkey Entry may cause grave consequences. More recently, Zhang \emph{et al.} \cite{zhang2019security} propose  downgrade attacks. Their insights are that the BLE specification does not enforce secure pairing for master devices such as a smartphone, leading insecure, one-way authentication. It can be observed that most of the attacks are possible due to the use of the insecure pairing protocol, such as Just Works \cite{4.2}, seeing that these insecure pairing protocols do not involve a trustworthy third-party or channel to hinder ``the man in the middle''. Our novel pairing protocol use the power supply charger and on-board clock to establish a trust channel, countering attacks innately.

%% file: Sections/sec3-pairing.tex
\section{IoT System and Pairing Protocols}
\label{sec:ble}
 
 In this Section, we will introduce the architecture of an IoT system, and then we discuss pairing protocols briefly. Particularly, we will provide more details of the pairing process of Bluetooth Low Energy, which is closely related to our work.  

\subsection{Architecture of an IoT system}

\begin{figure}  
\centering
\includegraphics[width=0.8\columnwidth]{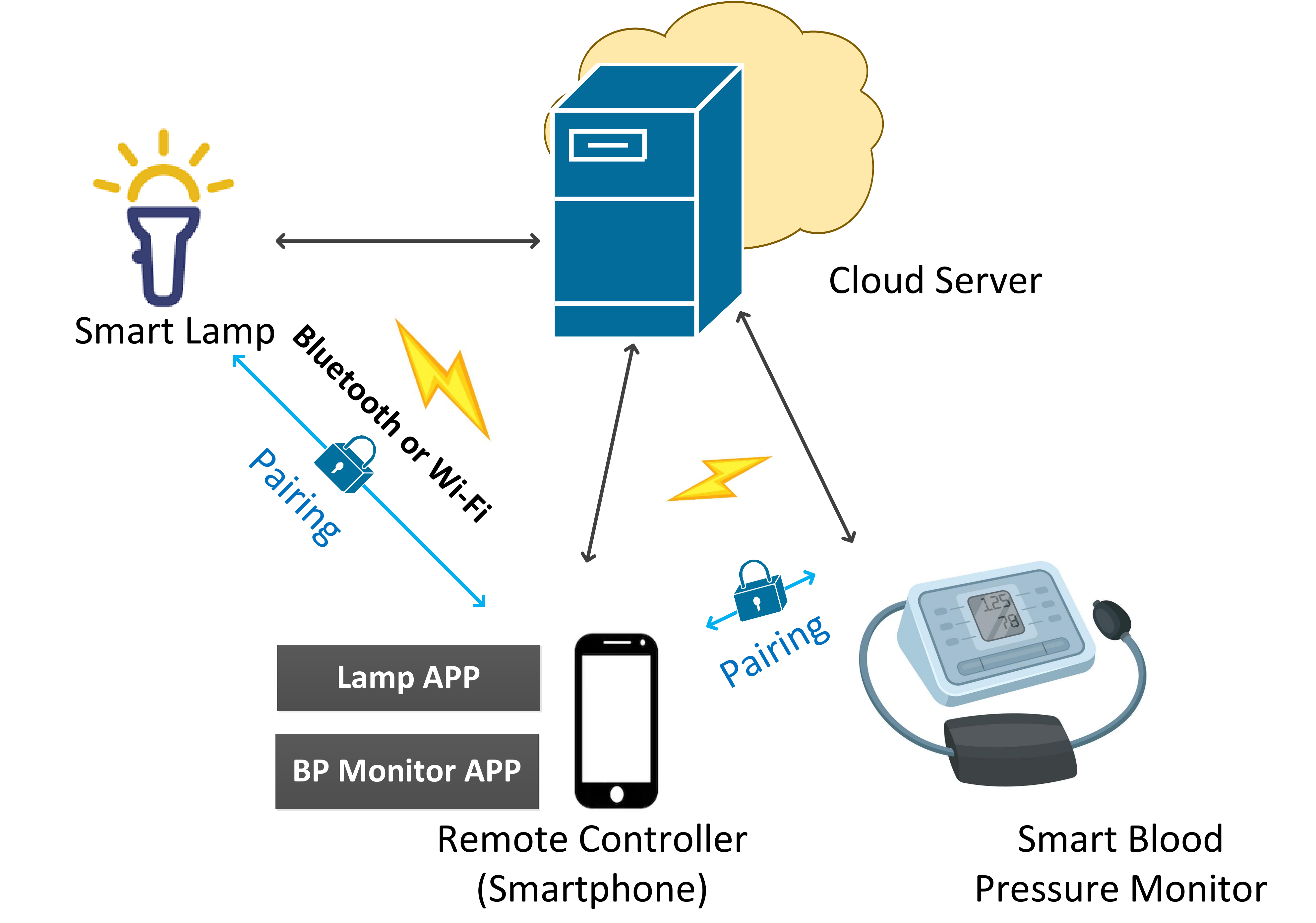}\\
\caption{Architecture of an IoT system}
\label{fig:iot}
\end{figure}

As demonstrated in Fig. \ref{fig:iot}, an IoT system usually involves a cloud server, some IoT devices and a remote controller such as a smartphone with specific apps. The IoT devices may implement various functions, such as medical monitoring or illumination. The remote controller is used to control and observe the status of the IoT devices. Basically, the remote controller and an IoT device  may communicate with each other directly through the same communication venue such as Bluetooth or Wi-Fi. However, when they are not in the same local network, the cloud server will relay the traffic between the device and controller. For a lighting system, the smart lamp is an IoT device, which provides the lighting service for users. The user may use his remote controller (e.g. his smartphone) to control the smart lamp through Bluetooth. 
Initially, a user may want to bind his controller with the IoT device via various communication venues, which is referred to pairing. Security is crucial factor for pairing process, considering that attackers may deploy eavesdropping or MITM attacks, which pokes vulnerabilities of IoT devices and compromises the entire IoT system.   

 \subsection{Pairing protocols}
 \label{subsec:workflow}

The pairing process is designed for devices that have security concerns. 
Basically, when pairing is enabled on two entities such as a remote controller and an IoT device, the same cryptographic key will be negotiated by both devices, such that the two devices can encrypt the communication using the negotiated key. To this end, devices may share the same static secret such as a pass-code or a key beforehand. However, these protocols are susceptible to reverse engineering. Attackers may extract these static secrets and deploy attacks directly.
Proximity-based device pairing schemes \cite{wu2017attack} are gaining popularity due to the fact that these protocols do not require a static secrets to be pre-shared, on condition that they can counter the reverse engineering innately.

As the term implies, proximity-based device pairing schemes require the two pairing devices and their genuine users are close to each other and the users are cable of operating the pairing devices freely, while it also assumes that the attacker is further away and does not have physical access to the pairing devices. Instead of sharing the secrets beforehand, the proximity-based device pairing schemes require the two devices exchange secrets in real time. To this end, trusted auxiliary channels such as audio-visual signal\cite{prasad2008efficient,saxena2006secure}, correlated human behavior\cite{chagnaadorj2013mimicgesture,4.2} may be involved.

 For the sake of easy presentation, we will elaborate on the pairing process of Bluetooth Low Energy and explain the details whenever needed. 
The pairing process of Bluetooth Low Energy consists of three phases: (i) the two devices exchange their pairing features. The pairing features here mainly refer to the input and output capabilities such as displays and keyboards. While other features, such as secure connections bits are also exchanged in this phase, we will not discuss since these features are out of our scope. (ii) Based on the exchanged features, the two devices then preform the authentication process and generate cryptographic keys. There are four authentication methods available for BLE, including ``Passkey Entry'', ``Numeric Comparison'', ``Out of Band (OOB)'' and  ``Just Works''. Among them, ``Out of Band (OOB)'',``Passkey Entry'' and ``Numeric Comparison'' are considered as secure pairing protocols, while ``Just Works'' is subject to Man-in-the-Middle attack. Secure pairing protocols are secure since these protocols involve  auxiliary channels to exchange secrets. (iii) After authentication, the two devices result in the same long term key. This key is used to derive session keys, which are used to encrypt links. When link is encrypted, other keys, such as an IRK (i.e., identity resolution key, which is used to preserve the privacy of Bluetooth devices) \cite{4.2} can be delivered from one device to the other. We will not discuss other keys due to the page limitation.  

%% file: Sections/sec4-switchpairing.tex
\section{SwitchPairing}
\label{sec:design}

In this Section, we first discuss the threat model. We present the high-level idea of our pairing protocol and explain the procedure afterward.  
 
 
\subsection{Threat model}

In our scenario, a user has one or more IoT devices, and he can control these devices via a remote controller. The ultimate goal of our protocol is to secure the communication of the pairing process. Particularly, we make the following assumptions: 

\begin{enumerate}
\item We assume the attacker does not have physical access to user's devices, including the IoT device, the remote controller and the power source which is used in our protocol. This is reasonable since these devices are sensitive gadgets, and people normally tend them carefully.
\item We assume that the pairing devices have power supply chargers and each of them has an onboard clock. Note that most of devices have power supply changers. Alternatively, if they do not have any (e.g., battery-based), vendors can enable the related interface to support one with negligible costs. For the onboard clocks, most of the devices/chips have electronic clocks \cite{grivet1963nonlinear}.
\item In some sophisticated attacks, the attacker may peep the operation of users, one difference to many other secure pairing protocols such as Passkey Entry of BLE is that, we do not make such an assumption. That is, previous protocols can not defend such a sophisticated attack.
\item We assume that the attacker can obtain the same type of device and the official app as the user does. This is reasonable since the apps are free to download, while an attacker can purchase the same type of device on the market with little effort. We made this assumption because the security of our protocol does not rely on the disclosure of the algorithm involved in our protocol. Attackers are free to know the workflow of our protocol.
\item We assume that the cryptographic algorithms are not the source of the vulnerabilities. For example,  attackers cannot break the algorithm such as SHA256 hash algorithm used in the protocol.  
\end{enumerate}

\subsection{Overview}

From Section \ref{sec:ble}, we know that the security of pairing depends on how the two devices exchange their messages for secret agreement. That is, if the secret is transferred from one device to another securely, the pairing protocol is secure. 
For example, in Passkey Entry of BLE, correlated human behavior is such a trusted channel. The owner of the two devices works as a relay, and he sees the six digits from one device and types it into another device.
Without loss of generality, we, therefore, define two channels in our paper: (i) Communication channel refers to the channel through which the data other than secrets will be transferred. (ii) Auxiliary channel refers to the trusted channel through which the secrets will be transferred. 
 
\textbf{Overall Solution}: Originally, for the devices that do not have any input/output capabilities, there is no other way except a communication channel for them to exchange parameters. Therefore, we may need to set up an auxiliary communication channel. Also, to keep our design economically viable, we must use these out-of-box components.

We use a power supply changer to set up the channel. Particularly, without connecting the two devices together via a power supply charger, which may subject to interface specification, we build an alternative communication channel via a power source (e.g. a plug).  Our idea is the following: (i) We connect two devices to the same power source, by using their own power supply charger. 
(ii) Most devices/chips have electronic clocks \cite{grivet1963nonlinear}. We let both devices synchronize clock frequencies via the communication channel. 
(iii) The owner of the two devices can then switch the power source on and off intentionally for a few times. If each device records the timestamps of the power switch off/on, the recorded timestamps contain enough amount of information, which can be used to derive the same encryption key. Also, the auxiliary channel used in our paper is secure enough to defend attacks such as eavesdropping attack and guessing attacks. In regards to this, we have a complete security analysis that proposed in Section \ref{subsec:analysis}.

 \begin{figure}  
 \centering
 \includegraphics[width= \columnwidth]{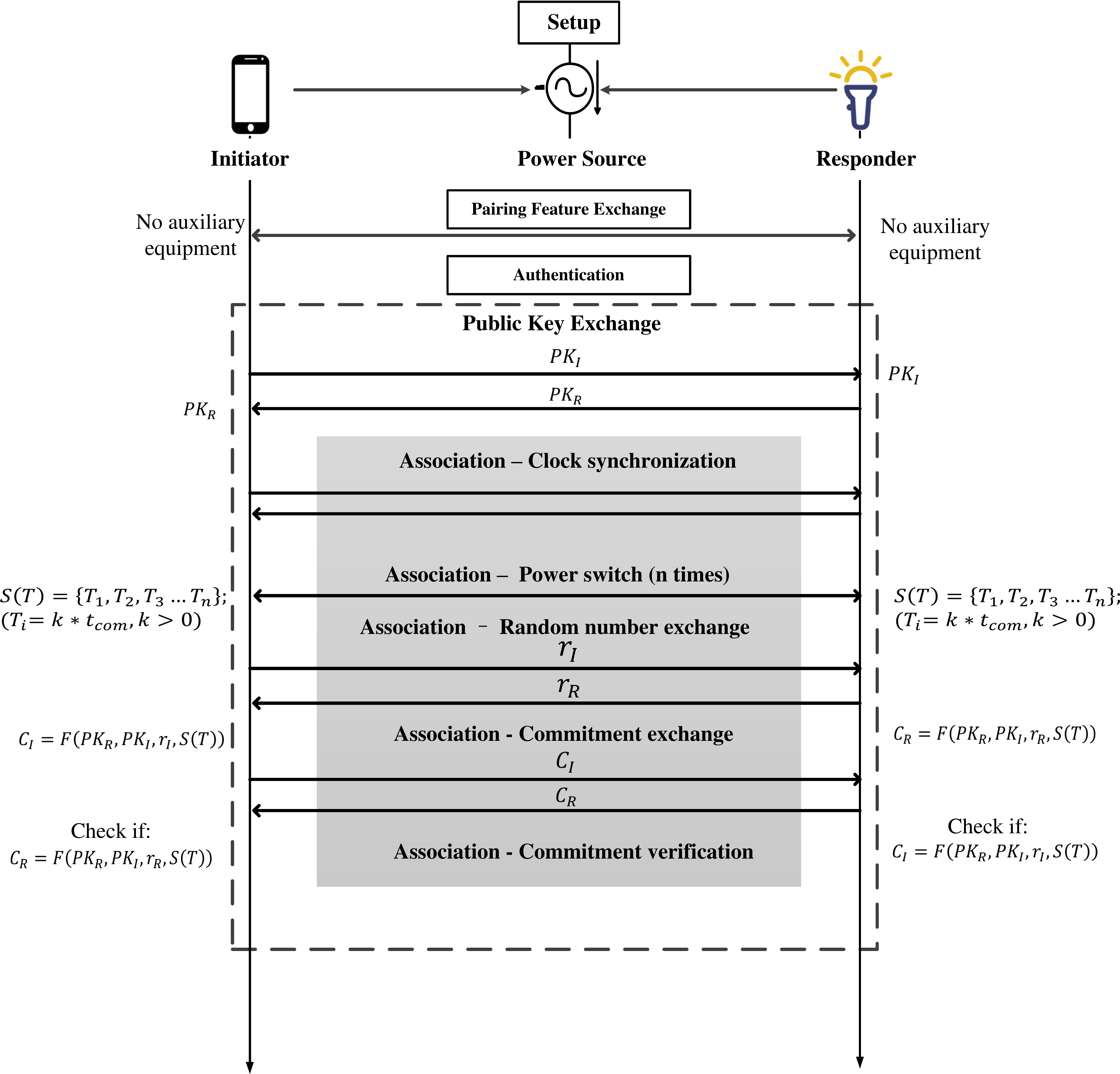}\\
 \caption{The workflow of SwitchPairing}
 \label{fig:swtichpairing}
 \end{figure}

  

 \subsection{Pairing procedure }

This Section will describe the pairing procedure of our design. For the sake of easy presentation, we denote the pairing initiator as $I$, and denote the pairing responder as $R$. We assume that each of the pairing devices has a public/private key pairs.   
We also assume that all the pairing devices do not have any auxiliary equipment.
As illustrated in figure \ref{fig:swtichpairing}, we elaborate on our pairing protocol below and notations to be used can be found in Table \ref{tab:notation} for the easy presentation:

\begin{table} 
\centering 
\caption{Summary of notations}
\label{tab:notation}
\begin{tabular}{|c|P{6cm}| }
 \hline
{\bf Notation} &  {\bf Description  }     \\
\hline
  $ PK_{i} $  &  the public key of a pairing device $i$ \\
 \hline
  $ SK_{i} $  &  the private key of a pairing device $i$  \\
 \hline
   $F(.)$  &  a hash function   \\
 \hline
   $H(.)$  &  a hash function other than $F(.)$ \\
 \hline
    $DH(.)$  &  the Elliptic-curve Diffie–Hellman (ECDH) key exchange algorithm \\
\hline
   $ P(.) $  &  a hash function  \\
\hline
\end{tabular}
 
\end{table}

\begin{enumerate}
\item \textbf{Setup}: Initially, the owner of the two devices plugs the two devices into the same power source. 
\item \textbf{Pairing Feature Exchange}: When the pairing process starts, $I$ and $R$ exchange their pairing features. At this moment, they are ready to initiate our SwitchPairing authentication. 
\item \textbf{Authentication - Public Key Exchange}: The two devices exchange public keys. We denote the exchanged public key as $PK_{I}$ and $PK_{R}$, where $PK_{I}$ refers to the public key of the initiator, while $PK_{R}$ refers to the public key of the responder.

\item \textbf{Authentication - Association}: This process is the core logic of our design. We list the steps below: 
\begin{enumerate}
\item  The two devices synchronize their clocks. 
Once synchronized, each clock starts timing simultaneously. We denote the initial timestamp as $T_{0}$. Please note that the initial time usually will not be a part of the secret. 
\item  The user switches off the power source, so the power supply of both devices goes off.  Both devices lose their power at the same moment $t_{1}$, which will be recorded individually. The user turns the plug on afterward.  The user repeats this step for $n-1$ times ($n>4$), depending on the security requirements of the application scenario, which will result in $n$ timestamps. At this time, each device has:
\begin{equation}
\begin{split}
 S(T)=\{t_{1},t_{2},t_{3}...t_{n}\}  
\end{split}
\end{equation}



It is worthy-noting that there is a delay between the time when a user switch-off/on the power source and the time when devices really are turned off/on. Such a delay is unavoidable due to the working principle of AC-DC converter \cite{lee2009advanced}. However, due to the diversity of integration and implementation, the delay may also vary from one device to another, causing the timing obtained on each device is different, which shall not be ignored.  Regarding this, we proposed our processing strategy to counter the delays. Please refer to Section \ref{subsec:timeprocess} for more details.   

\item Each device generates a random number $r_{I}$ and $r_{R}$ respectively. Then they exchange the random number.
\item  Each device feeds the random number and their public key together with $S(T)$ into a function $F(.)$  to compute a commitment. The initiator will have $C_{I}$:
\begin{equation}
\begin{split}
C_{I}=F(PK_{R},PK_{I},r_{I},S(T))
\end{split}
\end{equation}
and the responder will have $C_{R}$:
\begin{equation}
\begin{split}
C_{R}=F(PK_{R},PK_{I},r_{R},S(T))
\end{split}
\end{equation}
Each device exchanges the commitments. Then each device verifies if  random numbers, public keys and secrets can result in desired commitments. If yes, the same key can be generated from these parameters:

\begin{equation}
\begin{split}
DHKey =  DH(SK_{A},PK_{B})\\ =  DH(SK_{B},PK_{A}) 
\end{split}
\end{equation}

where $SK_{A}$ and $SK_{B}$ are the corresponding private keys of $PK_{A}$ and $PK_{B}$ respectively.

\begin{equation}
\begin{split}
Key = P\{DHKey,S(T),r_{R},r_{I}\}
\end{split}
\end{equation}

Please note that the demonstration is presented in the case of two pairing device, while our protocol can be also adopted on multiple device with little efforts. In the case of multiple devices, all devices shall be connected to the same power source and exchange public keys to each other (In the end of our pairing protocol, every two devices shall generate a key that used to communicate with each other), while the rest procedures are similar and we will not go to details due to the page limit. 

\end{enumerate}



\end{enumerate}
  
 \subsection{Delay processing}
 \label{subsec:timeprocess}
 
 Recall that there is a delay between the time when a user switch-off/on the power source and the time when devices really are turned off/on. This delay may vary from one device to another due to the manufacture. We introduce delay tolerance and fault tolerance to address this problem.

 \textbf{ Delay Tolerance}:
  We define the delay tolerance to handle this issue. With the delay tolerance defined,
 when a user preforms the switch-off operations, if the difference between two delays is less than delay tolerance, the two pairing devices can be considered as losing the power at the same time. Similar principles are also adopted when a user switches on the devices. For example, we assume the delay tolerance is $\tau$. When user performs the switch-off operation, the time point of losing power for the initiator is $t_{I}$, while that for the responder is $t_{R}$.  If the equation \ref{eq:delay} is satisfied, the two devices can be considered as losing power at the same time.   
 
 \begin{equation}
\begin{split}
|t_{I} - t_{R}| < \tau
\end{split}
\label{eq:delay}
\end{equation}

However, obtaining the difference between $t_{I}$ and $t_{R}$ is tricky since the timestamps can not be transferred through the communicating channel. Our idea is that we let $\tau$ equal the common time precision of the pairing devices (i.e., the minimal measurement of time for the pairing devices). In such a way, we can change $\tau$ by modifying the the time precision of the pairing devices without transferring the time points. The common time precision of two pairing devices is defined as follows:

 \begin{equation}
\begin{split}
\tau = tp_{com} = MAX\{ tp_{I}, tp_{R}\}  
\end{split}
\end{equation}

where  $tp_{I}$ and $tp_{R}$ are the time precision of the pairing devices, respectively. 
For multiple devices, a common delay tolerance shall be negotiated in a similar way. We assume that $m$ devices are connected to the same power source and when a user presses the switch off/on button of the power source, each device obtains the timing of losing power for itself, which can be represented as follows:

 \begin{equation}
\begin{split}
  T_{n}=\{t_{1},t_{2},t_{3}...t_{m}\}
 \end{split}
 \end{equation}

Therefore, we define the common delay tolerance for multiple devices as follows:

 \begin{equation}
\begin{split}
\tau_{com} = MAX\{ |t_{i}-t_{j}|\}  = MAX\{ tp_{i}, tp_{j}\} 
\end{split}
\end{equation}

where:

 \begin{equation}
\begin{split}
i \neq j ; 0<i,j\leq m ; m > 2;
\end{split}
\end{equation}

In Section \ref{sec:eva}, we explore the delay tolerance for two CC2640R2F chips to show the correctness of our principles. 
 
\textbf{ Fault tolerance}: Another factor that may mitigate the delay issue is the fault tolerance. The fault tolerance in our context is the capability that allows the pairing procedure to continue computing the key properly in the scenario where the timestamps obtained by each pairing device are not exactly identical. Besides, for the same operation performed by the user (i.e., a switch off/on operation), if the timestamps obtained by each device are different, this event is referred to an error. The number of errors over the number of switch off/on operations is denoted as the error rate. Basically, a lager fault tolerance will allow more errors. We introduce our algorithm to enable the devices compute the error rate before the key generation without exchanging timestamps. 

To this end, both device may want to exchange data other than the timestamps via the communicating channel and the data exchanged is termed evidence in our protocol. We first use the scenario where there are two pairing devices, denoted as device $A$ and device $B$, who want to pair up, and discuss the difference of that on multiple device afterwards. We assume that the user turns off/on the power source $n$ times ($n>4$), which will result in $n$ timestamps on each device. At this time,  device $A$ has:
\begin{equation}
\begin{split}
 S(T)^{A}=\{t^{A}_{1},t^{A}_{2},t^{A}_{3}...t^{A}_{n}\}  
\end{split}
\end{equation}

while device $B$ has:
\begin{equation}
\begin{split}
 S(T)^{B}=\{t^{B}_{1},t^{B}_{2},t^{B}_{3}...t^{B}_{n}\}  
\end{split}
\end{equation}

device $A$ and device $B$ then use the same hash function, denoted as $H(.)$, to generate a vector, named evidence, respectively, which can be represented below:

 \begin{equation}
\begin{split}
H(S(T)^{A})=\{H(t^{A}_{1}),H(t^{A}_{2}),H(t^{A}_{3})...H(t^{A}_{n})\}  
\end{split}
\end{equation}

 \begin{equation}
\begin{split}
H(S(T)^{B})=\{H(t^{B}_{1}),H(t^{B}_{2}),H(t^{B}_{3})...H(t^{B}_{n})\}  
\end{split}
\end{equation}

Afterwards, the two pairing device exchange their evidences and each device can compute the error rate $\varepsilon$ respectively:

\begin{equation}
\begin{split}
 \varepsilon = \frac{ \sum_{i=1}^n H(t^{B}_{i})\oplus  H(t^{A}_{i}) }{n}
  \end{split}
\end{equation}

where:

\begin{equation}
\begin{split}
 4 < i < n 
  \end{split}
\end{equation}

In such a way, if $t^{A}_{i} \neq t^{B}_{i}$, then $ H(t^{A}_{i}) \oplus H(t^{B}_{i}) =1$ due to the property of hash functions. Therefore, each device can know if there is any errors during the pairing process without exchanging the timestamps. Meanwhile, if and only if the error rate is less than fault tolerance, the two devices can continue the pairing process. Otherwise, the two pairing devices will tear down the connection and abort the pairing process. 
For multiple devices (e.g. $S(D)= {D_{1},D_{2},...D_{m}}$), each device will obtain a matrix after a user turns off/on the power source $n$ times ($n>4$), which can be represent as follows:
$$
 \left\{
 \begin{matrix}
   H(t^{D_{1}}_{1}) &  H(t^{D_{1}}_{2}) & ....& H(t^{D_{1}}_{n}) \\
   H(t^{D_{2}}_{1}) &  H(t^{D_{2}}_{2}) & ....& H(t^{D_{2}}_{n}) \\
   H(t^{D_{3}}_{1}) &  H(t^{D_{3}}_{2}) & ....& H(t^{D_{3}}_{n})\\
   ...&...&...&...\\
   H(t^{D_{m}}_{1}) &  H(t^{D_{m}}_{2}) & ....& H(t^{D_{m}}_{n})
  \end{matrix}
  \right\}  
$$
 Due to the property of hash function, if and only if the rank of this matrix equals 1, the pairing process contains no errors. Otherwise, the error rate $\varepsilon$ can be represented as follows:
 
 \begin{equation}
\begin{split}
 \varepsilon = \frac{ \sum_{i=1}^m H(t^{D_{1}}_{i})\oplus  H(t^{D_{2}}_{i}) \oplus ...H(t^{D_{n-1}}_{i}) \oplus H(t^{D_{n}}_{i}) }{m}
  \end{split}
\end{equation}
 
 where:

\begin{equation}
\begin{split}
 4 < i < m 
  \end{split}
\end{equation}
 


It is worthy-noting that there is a trade-off between the fault tolerance and security, developers may want to specify the fault tolerance carefully according to the context. For mission critical applications, a low fault tolerance should be enforced to hinder potential attacks.



%% file: Sections/sec6-implemetation.tex
\section{Implementation Criteria}
\label{sec:implementation}
\begin{figure}
     \centering 
     \begin{subfigure}[b]{0.5\textwidth}
         \centering
         \includegraphics[height=130pt]{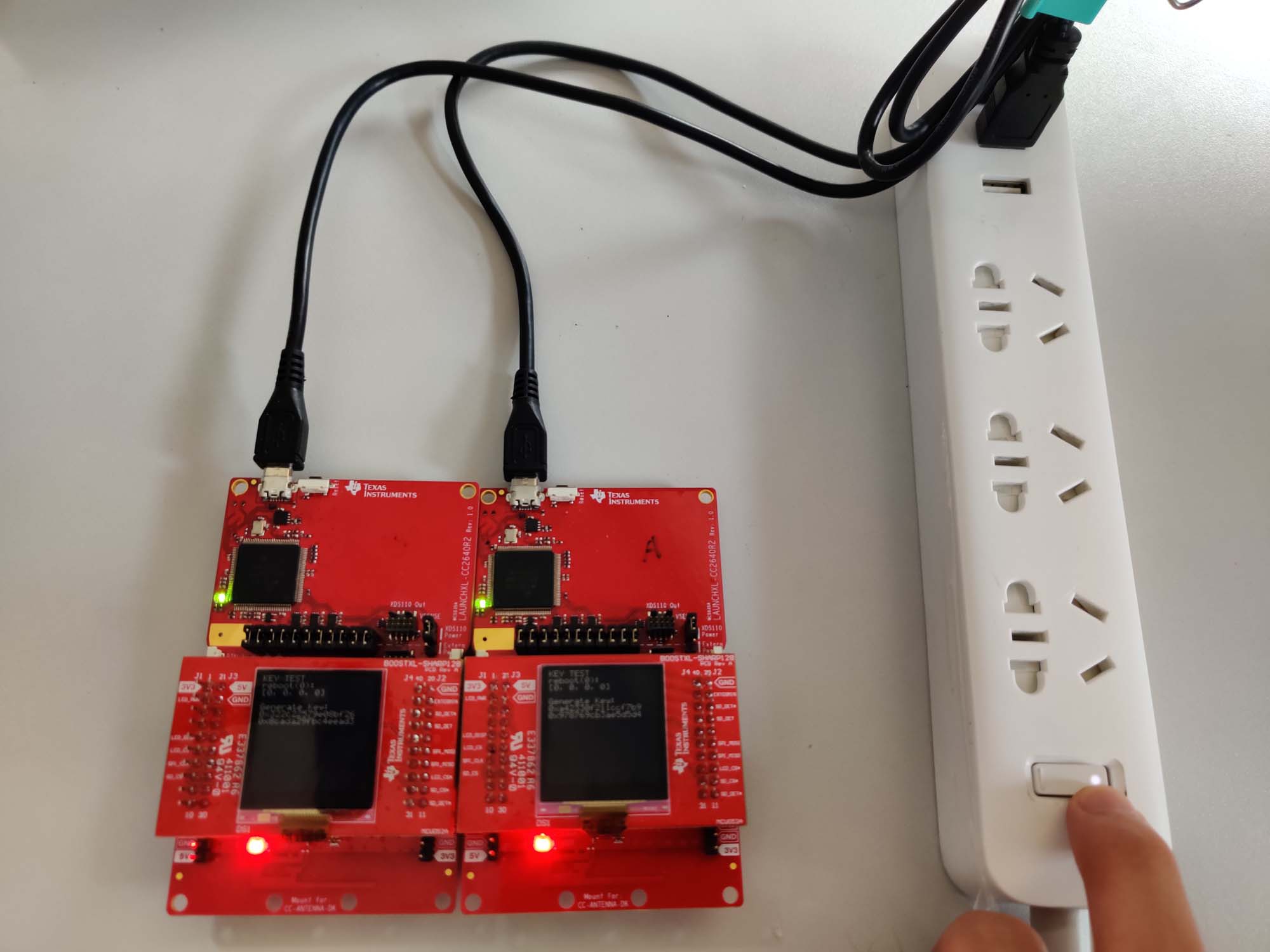}
         \caption{A user performs the association stage}
         \label{fig::prototype1}
     \end{subfigure}
     \vfill
     \vspace{2mm}
      \begin{subfigure}[b]{0.5\textwidth}
         \centering
         \includegraphics[height=130pt]{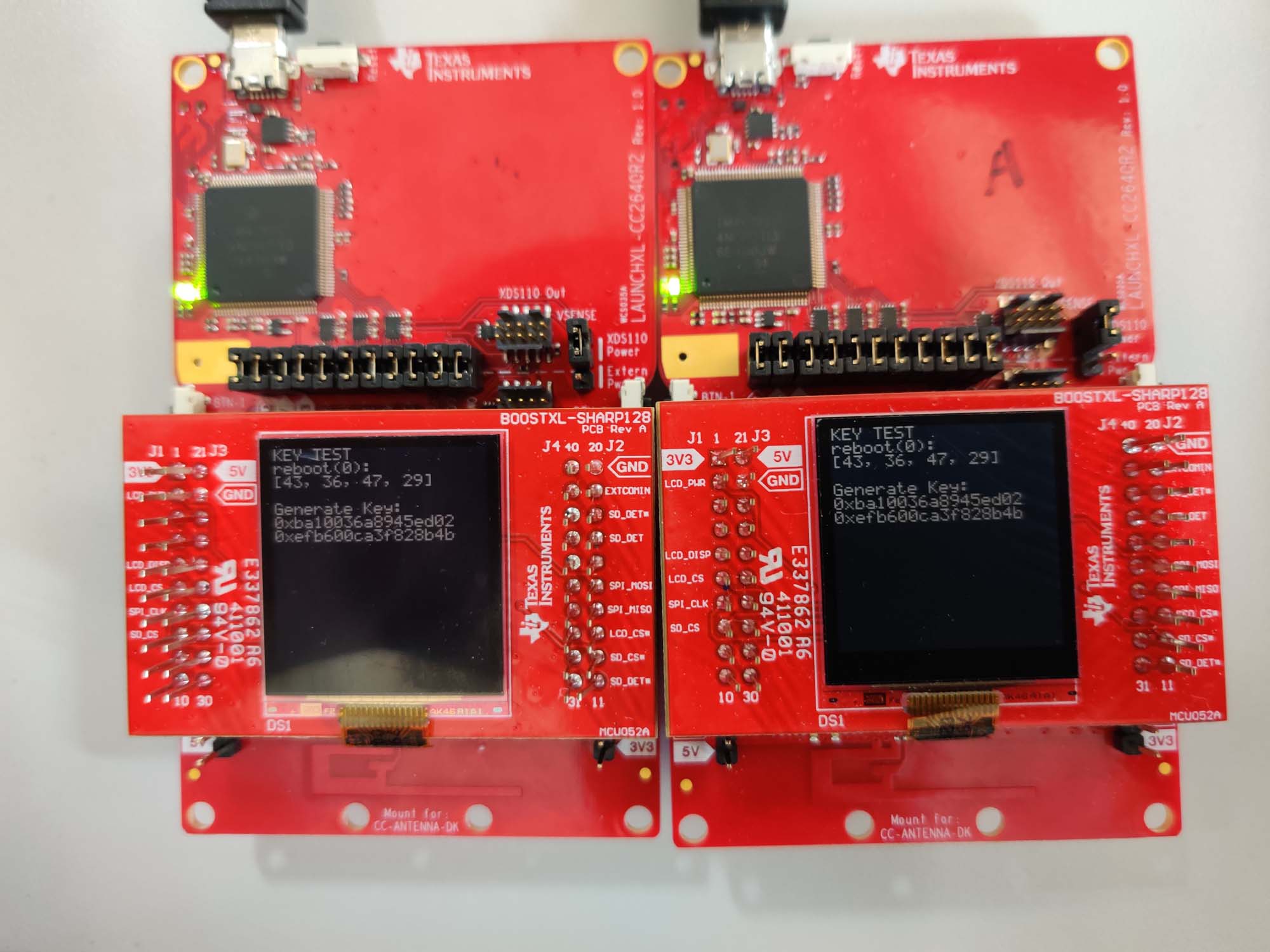}
         \caption{The two secrets displayed are the same}
         \label{fig::prototype2}
     \end{subfigure}
        \caption{The prototype of our SwitchPairing }
        \label{fig:prototype}
      
\end{figure}

We implement the proof-of-concept on two  CC2640R2F  development boards from Texas Instruments (TI) to show the feasibility of our SwitchPairing. 
Fig.~\ref{fig:prototype}  illustrates the prototype. In Fig.~\ref{fig::prototype1}, a user operates the plug to perform the association stage. In Fig.~\ref{fig::prototype2}, the two devices generate the same secret. Please note that our solution suits all kinds of devices, even those without any input/output components. For the convenience of the demonstration, we install LCD display modules on both development boards. For example, by comparing the timestamp sequences, we can determine whether both devices have shared the same secret.

 
The SDK platform from TI creates a variety of developing solutions for developers to build their own products. In regards of security, TI CC2640R2F has an AES accelerator and an ECC library in ROM. Therefore, we may use the off-the-shelf cryptographic APIs, without having to reimplement the algorithms. In particular, our protocol is built on the top of a typical BLE peripheral solution named ``Simple Peripheral'', since this solution integrates BLE stack and a simple BLE application that handles the basic communication process. 

To implement the protocol, we have a few major concerns. First, our protocol involves the clock synchronization and beating mechanism, and these should be carefully crafted. Second, keys should remain in the flash of the chip when the power goes off so that the pairing process will not repeat. To address the above issues, we have the following solutions.  

\textbf{Clock mechanism:} In TI's SDK, clock instances are used to define delayed or periodic tasks. The TI's  Real-Time Operating System (TI-RTOS), which is a light-weight, multi-thread operating system, will schedule such tasks according to a given number of systems ticks. We use clock instances to design a heartbeat mechanism. Specifically, a device will start the synchronization with its peer device when it is in the pairing stage. Meanwhile, a clock instance is initialized and starts to perform a periodic heartbeat task every once in a cycle period of $T$. We denote this $T$ as temporal precision for the easy of presentation. Particularly, according to our experiment, the optimal temporal precision for the TI CC2640 is 50 ms. That is, in every second, the TI CC2640 may generate 1000/50=20 different values that can be used to derive the key. In such a way, we can track the moment when users switch on/off the devices and generate a secret that is secure enough.

\textbf{Secret Key storage:} The Simple NV (SNV) is an area in the flash memory, which is dedicated for persistent data. According to the document, the data stored in this area will not lost when power is off; it can be used to store sensitive data such as encryption keys. Therefore, we store the heartbeat data in this flash memory area with provided APIs \textit{osal\_snv\_read} and \textit{osal\_snv\_write}. Moreover, to protect hardware-based cracking, we disable the debugging
ports after we finish the implementation.


    
    
    
    
The implementation criteria offer the community good guidelines to follow. 
Note that our criteria can also be extended to other chips and other communicating venues with little efforts. We use the TI CC2640 as a demonstration due to its prevalence. Particularly, periodic tasks and persistent memory are two fundamental features to craft our protocol. Therefore, vendors can always implement our protocol on other chips with such features enabled.

%% file: Sections/sec5-secanalysis.tex
\section{Security Analysis}
\label{subsec:analysis}
The pairing procedure of our design follows the procedure of standard pairing protocol (e.g, Passkey Entry) strictly so that our design will not downgrade the security of secure pairing protocols. Particularly, our SwitchPairing defends the passive/active eavesdropping attacks, as well as guessing attack without any changes.

\subsection{Passive eavesdropping attack}

To deploy the passive eavesdropping attack in our context refers to the attacker can obtain the actual timestamps that the user presses the button of power source passively, by using some kind of devices such as a sniffer. After obtaining timestamps, the attacker then uses the timestamp to compute the secrets. However, the secret is transferred through our ``auxiliary channel''. As discussed in our threat model, it is impossible for a regular attacker to deploy such an attack because they do not have physical access to user's personal devices such as an IoT device or a power source, on condition that the attacker has no chance to bug the power source.

\subsection{Active eavesdropping attack}

Deploying active eavesdropping attack in our context refers to such a scenario where a strong attacker can supervise the user somehow, such as an attacker stands around the user and peeks the user's operations. Many legacy pairing protocols,  such as PIN based pairing protocols or Visual/Audio-based pairing protocols, can not defend such a type of attack.  For example, the PIN based pairing protocol will fail if the attacker can peek the six digits PIN. In such a case, the attacker may block the victim devices and deploy the MITM attack with no change.

Our protocol can defend this type of attack. This is because the reaction speed of a human subject is relatively slow when compared to machines. Even if the attacker observes that a user presses the button, it still takes a while for him to handle the information and record the actual time. Regarding this, we have a complete evaluation in Section \ref{subsec:simattack}. 

\subsection{Guessing attack}

In guessing attack, instead of passively or actively eavesdropping the secrets,  the attacker may want to obtain the correct secret by guessing. The success rate of defending the guessing attack depends on the amount of information contained in the transferred secret. Therefore, the exchanged parameter should be strong enough to hinder the guessing attacks. For example, in Passkey Entry protocol, the exchanged secret is set to a 6-digits number, ranging from ``000000'' to ``999999''. There are 1000000 possibilities. Given that the attacker only has one chance to make the secret right, the success rate for an attacker deploys such an attack is 0.000001. 

Our pairing protocol can also defend the guessing attack effectively. Particularly, we assume the common delay tolerance is $t_{com}$ ms, so that it can generate $1000/t_{com}$ different values in a second. We assume that during the pairing process, the user performs $n$ time of switch off/on operations.
However, there is a time period between each timestamp due to the reaction time of the user. We assume the reaction time of a user is $T_{r}$ seconds. 
Therefore, we have the following equation, where $P$ is the probability that an attacker deploys the guessing attack successfully:  
\begin{equation}
\begin{split}
   P= \frac{1}{ (1000 \times T_{r}/t_{com})^{n} }  \label{eq:4}
\end{split}
\end{equation}

For the CC2640 chips used in our experiments (please refer to Section \ref{sec:implementation} and Section \ref{sec:eva} for more details), we have already known that the delay tolerance is $120$. We let $T_{r}=8$ and $n=4$, which refers to a user may complete the association process by performing the switch on/off 4 times, and the reaction time for this user is $8$ seconds. In such a condition, $P=\frac{1}{(\frac{(1000 \times 8)}{120})^{4}}\approx \frac{1}{18,974,736} = 5.3 \times 10^{-8}$. Compared with that of the Passkey Entry of BLE Pairing, which is $P=\frac{1}{999,999} = 1 \times 10^{-6} $, our Switchpairing is much more secure. Moreover, our design can be more secure when the user extend the association period or preforms the switch off/on operations for more time.   

\subsection{Active eavesdropping and guessing attack}

 A strong attacker may attempt to combine the active eavesdropping attack and the guessing attack together to craft a sophisticated attack. In such a scenario, an attacker is capable of supervising the user, and he may obtain an estimated time at which the user presses the buttons by barely observing the user's operations via his naked eyes. Therefore, he may guess the actual time based on the estimated time. We argue that this is also not feasible due to the attacker can only have one try. If the guessing attempt fails, the two pairing devices will result in the same key and will not repeat the pairing process anymore. Actually, when performing the active eavesdropping attack, the attacker is deploying this type of attack subconsciously. Thus, this sophisticated attack is subject to active eavesdropping attack and our pairing protocol can defend this without any changes. 

%% file: Sections/sec7-evaluation.tex
\section{Evaluation} 
 \label{sec:eva}
 
 In this Section, we conduct a set of evaluations to show the feasibility and security of our SwtichPairing protocol. Notably, our evaluation will cover the following aspects. We first show the impact of delay tolerance and switch off/on numbers.  Afterward, we measure the overhead of our SwtichPairing protocol.
 We then show how does SwitchPairing mitigate    various attacks during pairing process.
 Finally, we also demonstrate that our protocol runs on multiple devices properly. Please note that fault tolerance is configured to zero by default unless explicitly stated otherwise, which will provide a higher security level for our pairing process. However, in the case of SwitchPairing on multiple device, we also show the impact of fault tolerance. 
 
 \subsection{The impact of delay tolerance}
 
\begin{figure}  
\centering
\includegraphics[width= 0.9\columnwidth]{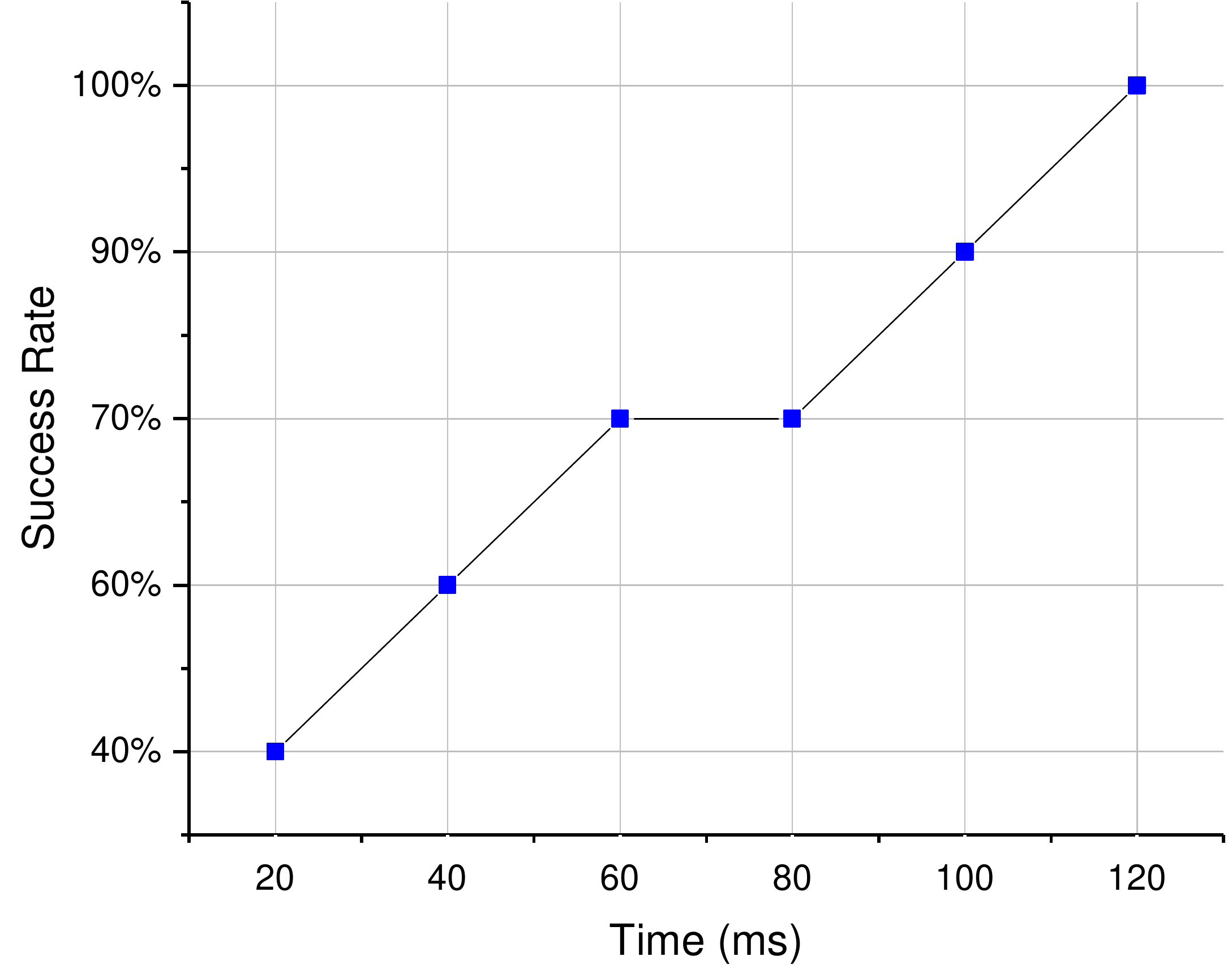}\\
\caption{The success rate vs. delay tolerance (two devices)}
\label{fig:srvstp}
\end{figure}

 Obviously, for two given devices, the security of our pairing protocol partly depends on their common delay tolerance. In our experiment, we use two CC2640 chips, and the delay tolerance should be considered as the same. This experiment is used to explore the common delay tolerance of TI's CC2640 chip. The common delay tolerance is set to 20, 40, 60, 80, 100, 120 milliseconds, respectively. 
 We perform the switch off/on operation 4 times as a test, and for each common delay tolerance, we carry out 10 tests; hence, the success rate is defined as the number of successful key generation (i.e., keys generated on both side are identical) over 10.
 Fig.~\ref{fig:srvstp} shows the result. It can be observed that when the delay tolerance is 120 milliseconds, the two devices can always generate the same key.
 
 
 \subsection{ The impact of switch off/on numbers }
 
\begin{figure}  
\centering
\includegraphics[width= 0.5\columnwidth]{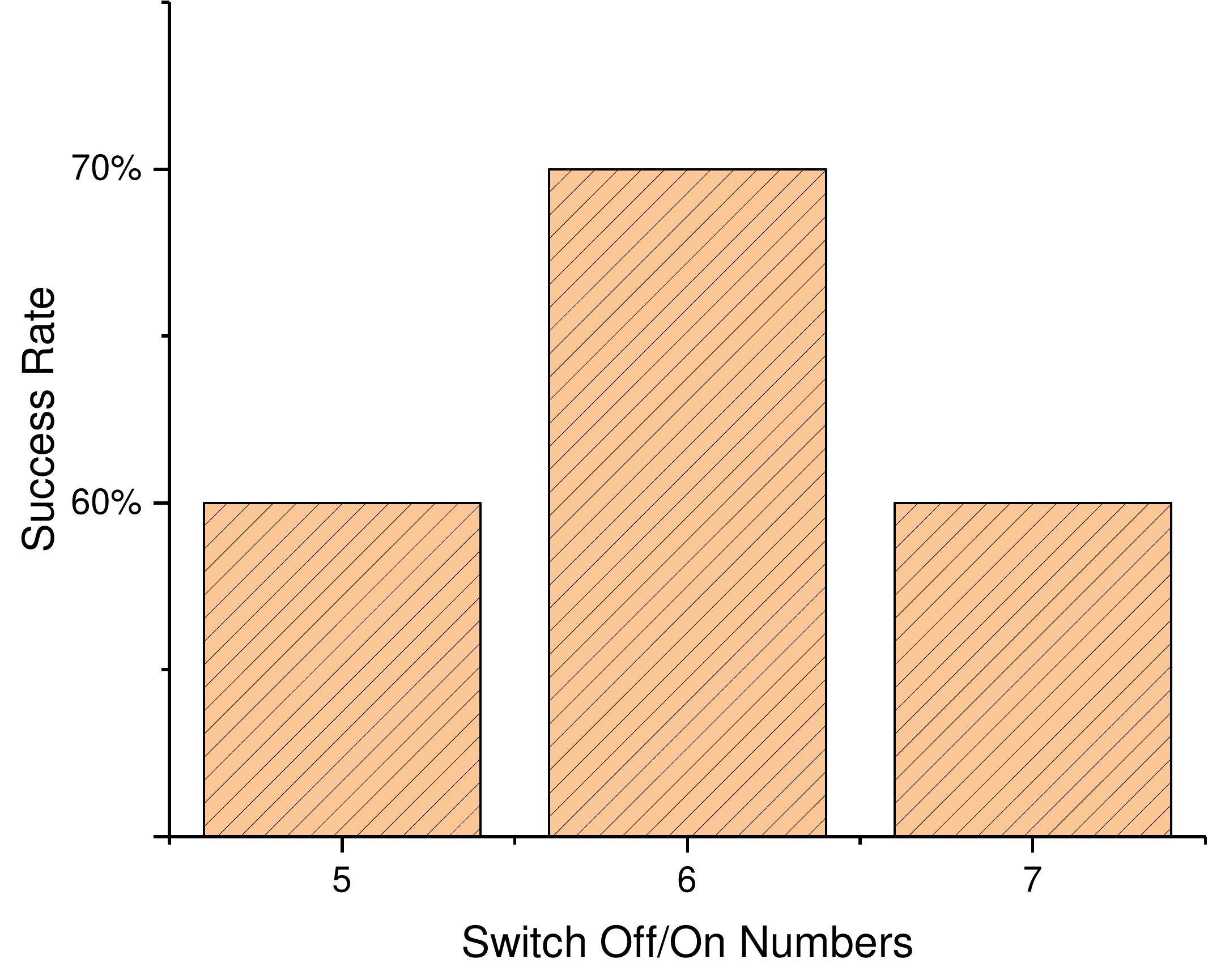}\\
\caption{The success rate vs. the number of switch off/on operations}
\label{fig:srvsnb}
\end{figure}

In addition to the delay tolerance, vendors may also increase the number of switch off/on operations $n$ to enlarge the keyspace, achieving stronger security protection, which we have elaborated on in our security analysis.  For instance, performing switch off/on four times to generate a key should be secure enough for home appliances, but it may fail to achieve the requirement of critical devices such as medical equipment. In this experiment, we will explore the impact of switch off/on numbers. To this end, we set the delay tolerance as 120 ms and perform switch off/on operations for 5, 6, 7 times, respectively. 
For each case, we carry out 10 tests; hence, the success rate is defined as the number of successful key generation (i.e., keys generated on both side are identical) over 10. 
Fig.~\ref{fig:srvsnb} shows the results.  We can observe that even when $n=7$, we can still get an acceptable, successful pairing rate. However, according to equation \ref{eq:4}, as $n$ increases, the probability $P$ of executing a guessing attack successfully will decrease exponentially. 

Some failures of the pairing process due to the issues of the device itself such as reboot failures.  This is why the success rate is lager when $n=6$ than that when $n=5$.  For the sake of rigor, we also factor in these results. We argue that these failures will be eliminated if the chips are better designed. For now, the vendors may want to tweak their delay tolerance to avoid this type of failure.

 
  \subsection{The overhead of our protocol}
 
\begin{figure}  
\centering
\includegraphics[width= 0.75\columnwidth]{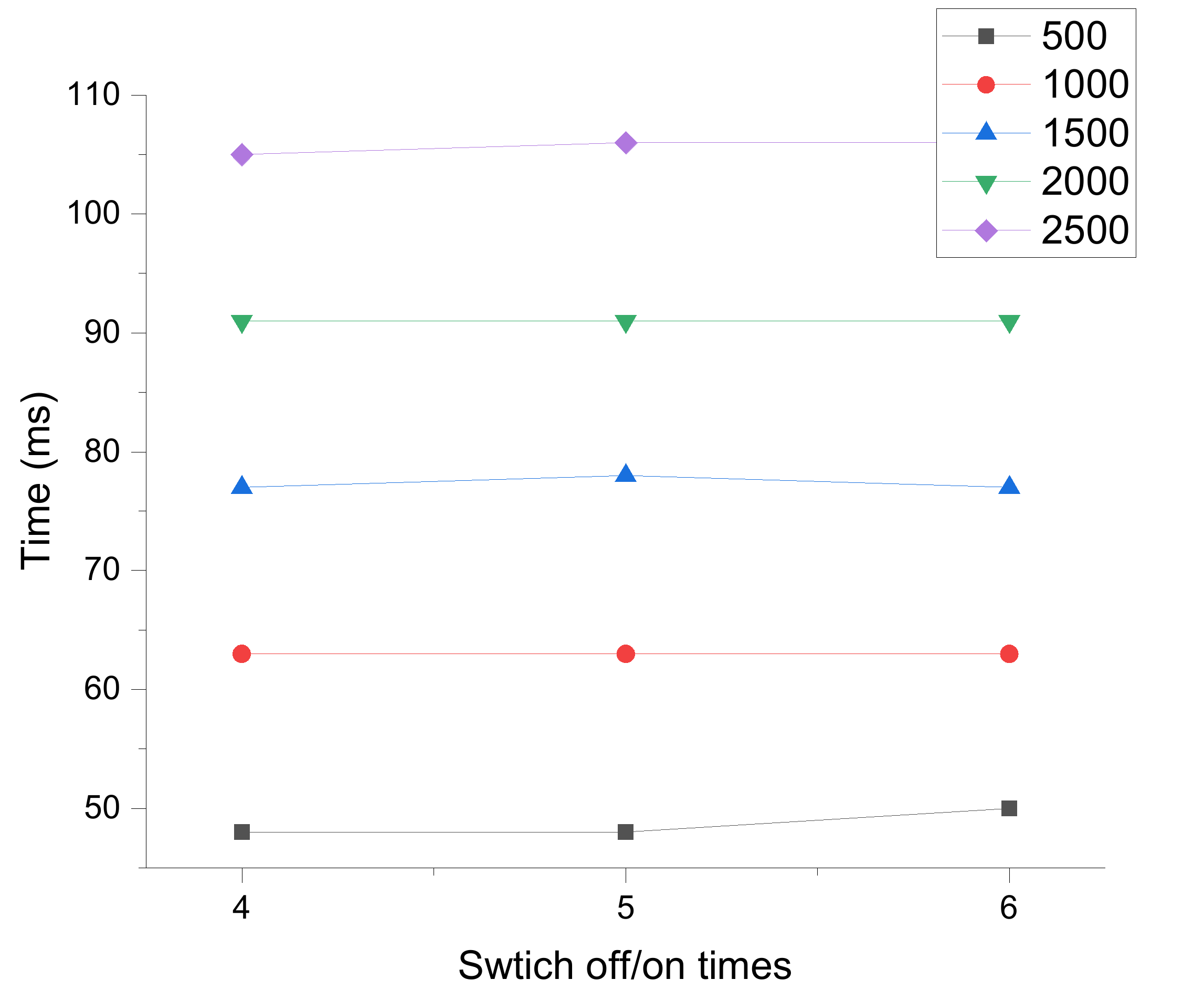}\\
\caption{ The overhead when MD5 is used}
\label{fig:md5}
 
\end{figure}

\begin{figure}  
\centering
\includegraphics[width= 0.9\columnwidth]{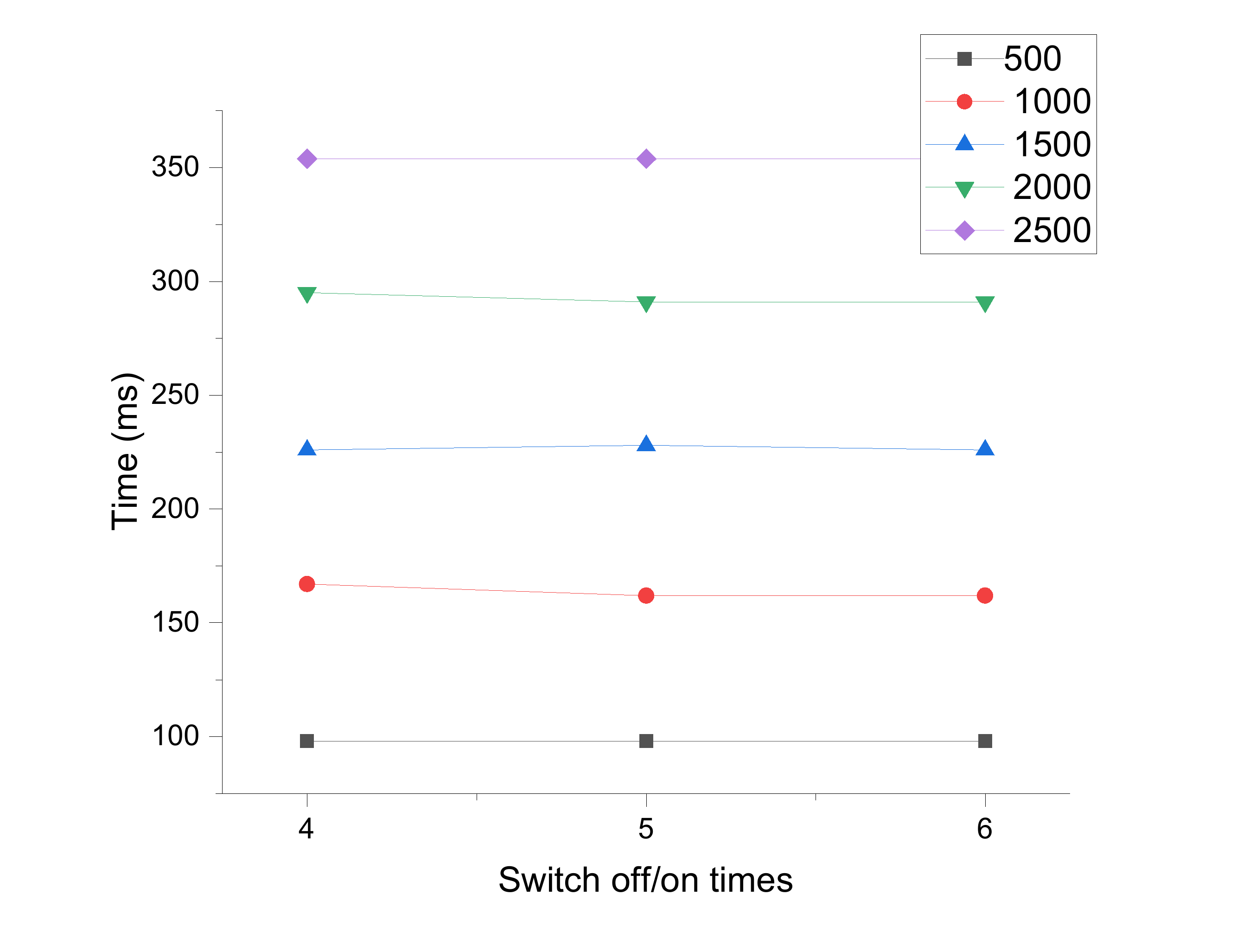}\\
\caption{ The overhead when SHA256 is used}
\label{fig:sha}
 
\end{figure}

  Recall that the secrets are exchanged directly in some pairing protocols such as Passkey Entry of BLE. However, in our protocol, the secrets is hashed from the timestamps. Therefore, we want to evaluate the overhead of our protocol, i.e., the overhead caused by the calculation of the secrets. Two types of hash algorithms are evaluated, including MD5 and SHA256. We run each algorithm 500, 1000, 1500, 2000, 2500 times accordingly and the input of the hash function is the timestamps. Particularly, we randomly generate 4, 5, 6 timestamps respectively. Fig.~\ref{fig:md5} and Fig.~\ref{fig:sha} give the results. Please note that we are aware of that the MD5 is not secure. We use MD5 as an example to demonstrate that vendors can achieve high security by using the SHA256 instead of using MD5, since with the SHA256 enforced, only little overhead is involved, as shown in the figures. It can also be observed that the overhead will not increase rapidly when we increase the number of preforming the switch on/off. Regarding this, we can enhance the security of our protocol by simply increasing the required switch on/off times.

 \subsection{ User Study I: The time intervals between operations}
  
 When a user operates the plug (i.e., switch on/off), we may want to know the time intervals between two operations. This is a fundamental parameter, which determines the other parameters related to the security of our protocol. As analyzed in Section \ref{subsec:analysis}, as the interval increases, the security of our protocol will increase rapidly. However, the operation speeds vary from one person to another. Therefore, we need to conduct a user study to set up a benchmark.
 Particularly, we are able to recruit 20 human subjects. All the human subjects are required to switch on and off the plug 5 times. We set the delay tolerance as 120ms and record the numbers of ticks between every two operations.  It can be observed that the numbers of ticks between every two operations is around 69 (i.e., $\frac{64.6   +   70.2   +    67.95   +   70   +   70}{5}=68.55$). Therefore, the time interval is around 69*120/1000 = 8 seconds.  
 
 
 \begin{figure}  
\centering
\includegraphics[width= 0.9\columnwidth]{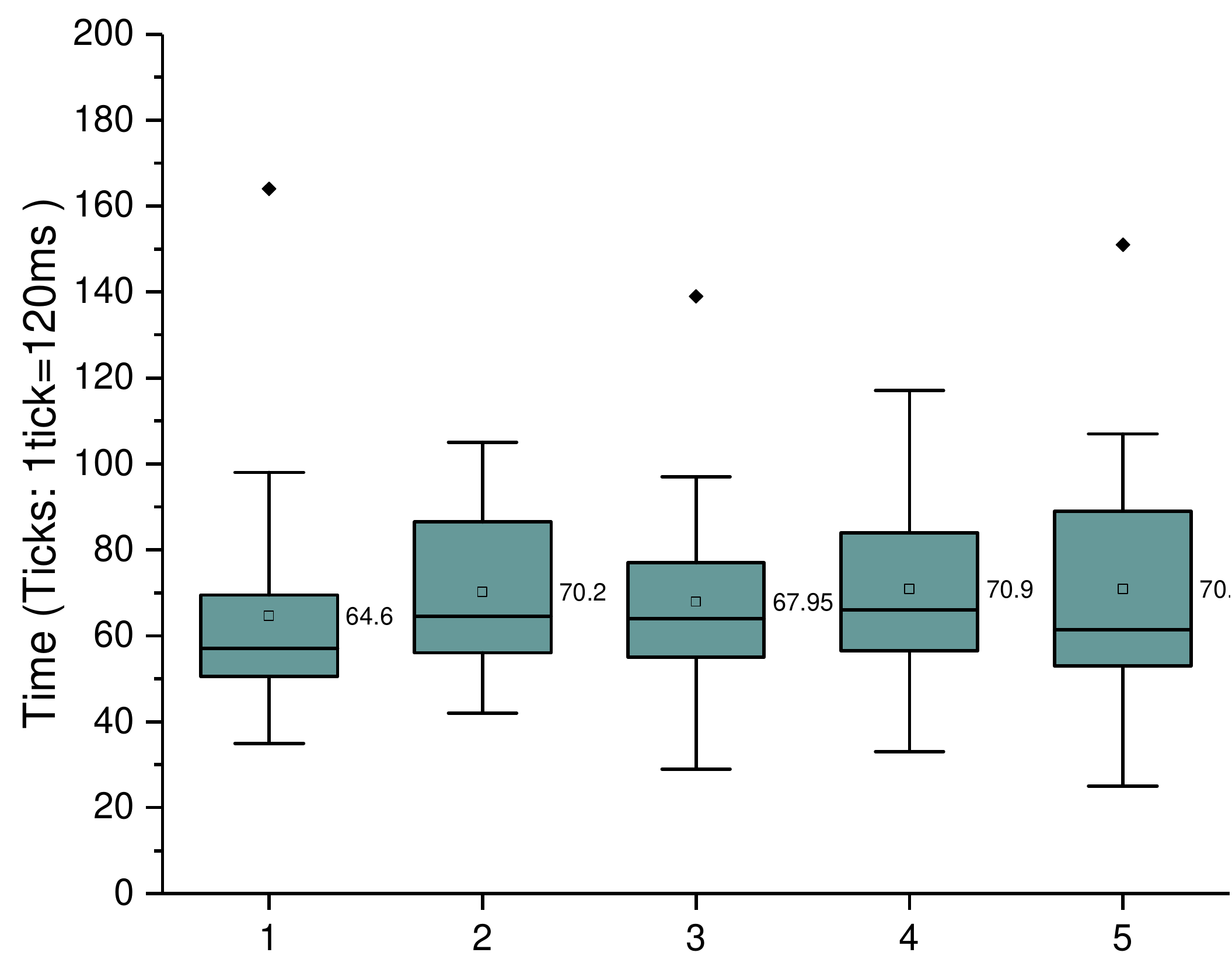}\\
\caption{The time intervals between operations}
\label{fig:boxplot}
\end{figure}

\subsection{ User Study II: The impact of power sources}
  
 This experiment explores the impact of different power sources. We conduct the experiment as follows. First, we plug the two peer devices on two different power sources (two plugs). Then, a volunteer is asked to perform the pairing process. Specifically, each hand of the volunteer is supposed to control a specific power source. For example, the left hand may control the plug, which connects a master device, while the right hand may control the plug with a salve device connected.  To conduct the pairing process, the volunteer may need to press the switch ``on/off'' button several times.  Particularly, we are able to recruit 5 human subjects as volunteers.  Before our experiment, we require the volunteers to press switch "on/off" button on each power plug synchronously every time.  We set the delay tolerance as 120 ms and the switch ``on/off'' time as four times. We run the test 10 times. 
  Table \ref{tbl:powersource} shows the results. It can be observed that the success rate is relatively low when they perform the switch off/on operations on different power sources. This experiment proves that the power source also has impact on the results. Therefore, for an attacker whose device is not connected to the same plug as the victim does, it is extremely challenging for the attacker to deploy attacks successfully.

 \begin{table} 
\centering
\caption{The impact of power sources}
\label{tbl:powersource}
\begin{tabular}{ cccc }
\hline
\textbf{User \#}  & \textbf{ Total Attempts } &   \textbf{ Succeed Attempts }& \textbf{ Success Rate} \\ \hline
\rowcolor{mygray}
1 &    10 & 0 & 0\% \\
 
2 &     10 &    1 &    10\%  \\
\rowcolor{mygray}
3 &  10  & 0 & 0\% \\
 
4 &     10 & 1 &    10\%  \\
\rowcolor{mygray}
5 &    10 &    2 &    20\%  \\
   
\hline
\vspace{1mm}
\end{tabular}
 
\end{table} 
  
 \subsection{User Study III: Simulated attacks }
 \label{subsec:simattack}
 
Passkey Entry is subject to an eavesdropping attack where a strong attacker can supervise the user somehow, such as the attacker stands around the user and peeps the Passkey on the display of the victim device. Once the attacker sees the Passkey, he may deploy the MITM attack. However, our protocol can defend such an attacker innately since the reaction speed of a human subject is relatively slow when compares to machines. According to a report from \cite{humanreactiontest}, the median reaction time of a human subject is around 215 ms, which is well above the bar used in our experiments (120 ms). To explore the feasibility of such an attack, we developed User Studies via 21 human subjects.
 
\begin{figure}  
\centering
\includegraphics[width= 0.9\columnwidth]{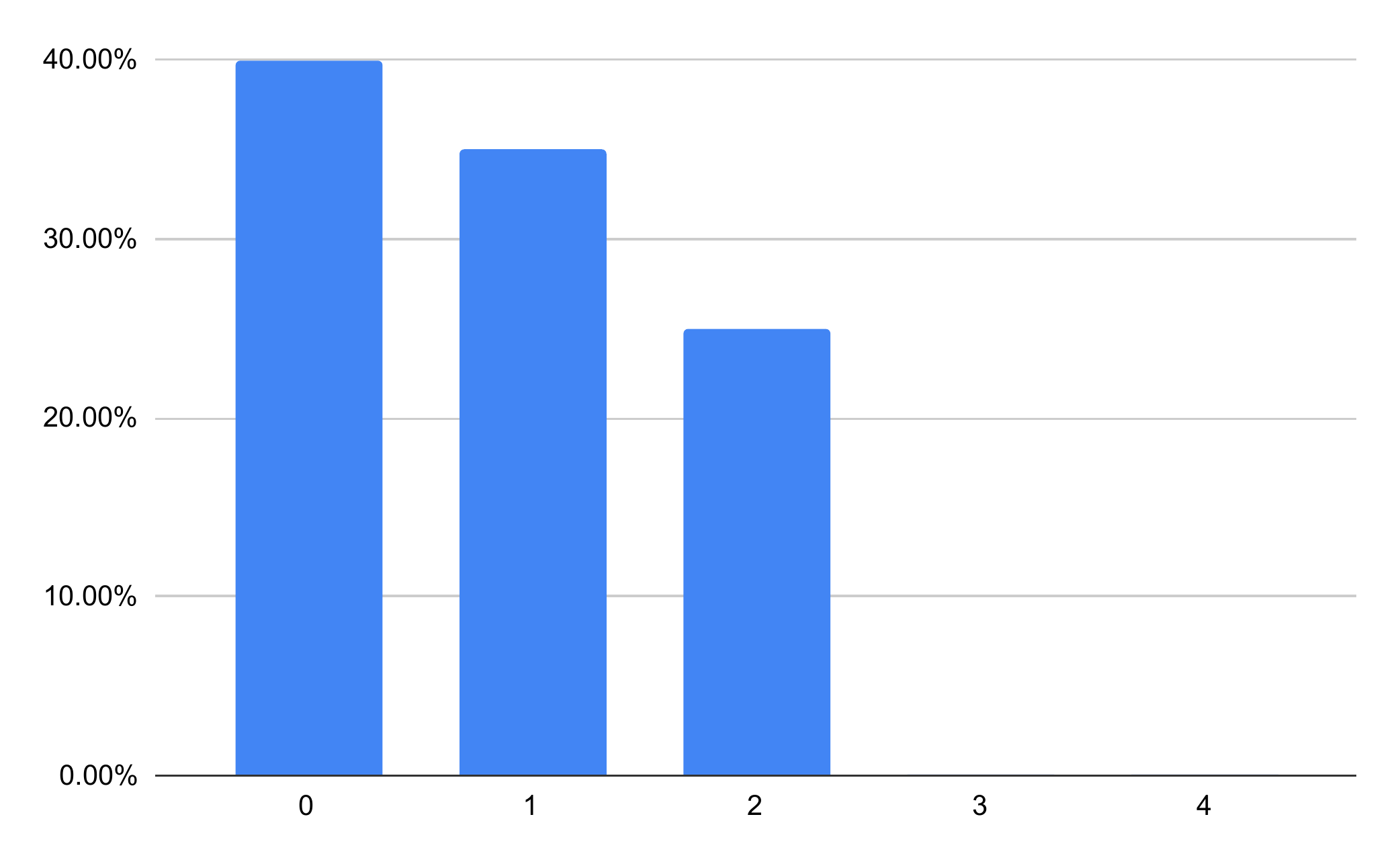}\\
\caption{ The number of success attacking attempts v.s. the number}
\label{fig:bt}
 
\end{figure}

 
 In the first experiment, the 20 human subjects pretend to be attackers, whose goal are to deploy the attack introduced above. We offer each human subject (i.e., the ``attacker'') a suite of our testing device: a power source with a development board plugged in. The ``attacker'' can switch off/on the power source freely by pressing the ``on/off'' button on the plug, which will lead the development board to reboot. The remaining human subject is asked to act like the victim. Basically, during our experiment, the victim will press the button 4 times.  
 According to our analysis in Section \ref{subsec:analysis}, a user switching on/off 4 times is secure enough when compared with the Passkey Entry of BLE. The ``attackers'' now are asked to 
 press the ``on/off'' button as soon as he/her sees the victim does so. We denote this action as an attacking attempt.  Fig.~\ref{fig:bt} shows the results, which demonstrates the relationship between successful attacking attempts (all 4 times) and the number of ``attackers''. For example, the first column indicates that there are 40\% attackers (which is 20 $\times$ 40\%=8), failing in all 4 attacking attempts, while the second column indicates that there are 5 attackers successfully deploying one attacking attempts.  Recall that if and only if all attacking attempts succeed, the same key can then be generated. Therefore, we conclude that there is no attacker succeed. However, in order to counter persistent attackers for higher security level, the number of switching shall be increased. 
  
 

 \subsection{SwitchPairing on multiple devices}
 
\begin{figure}  
\centering
\includegraphics[width= 0.8\columnwidth]{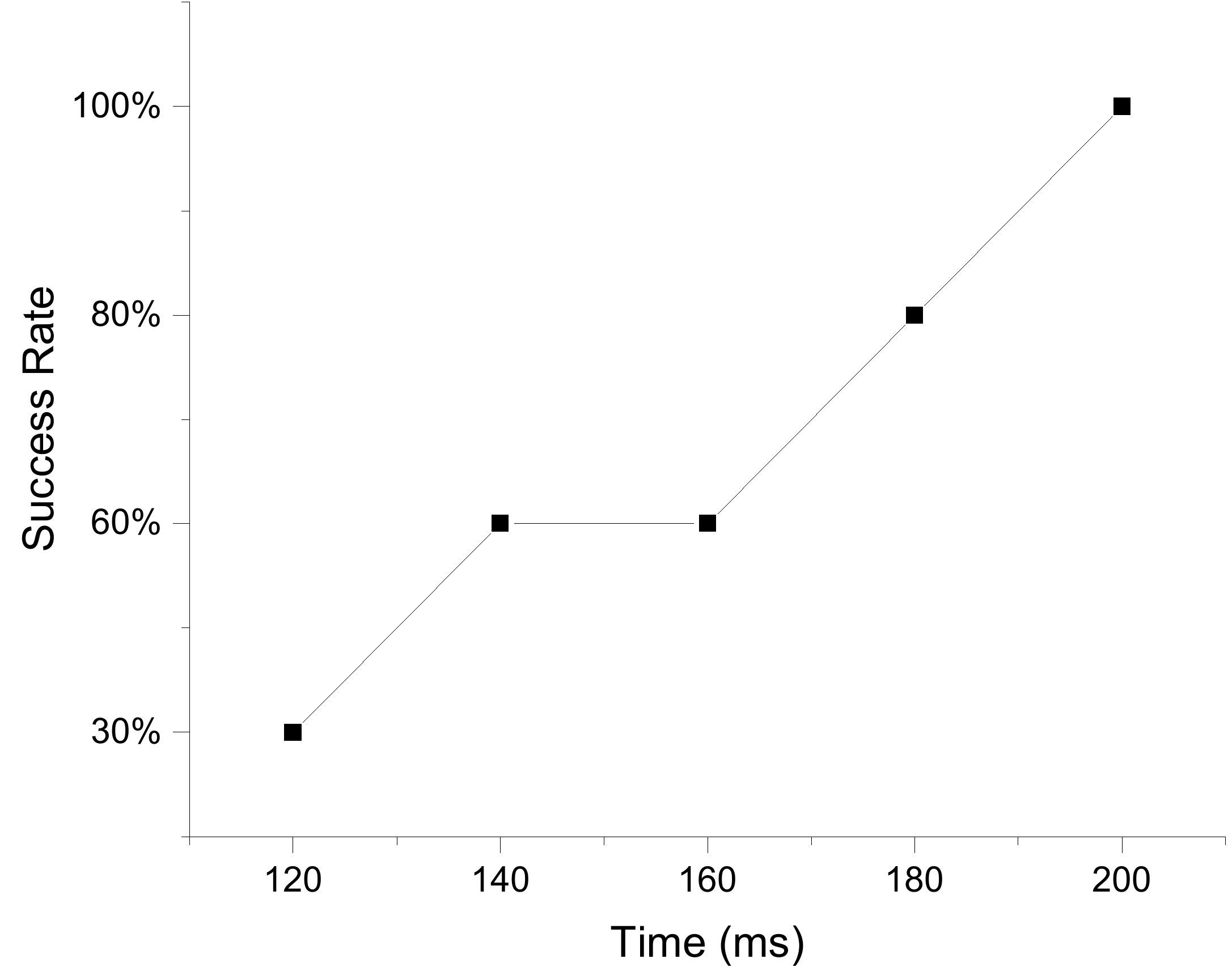}\\
\caption{The success rate vs. delay tolerance (multiple devices)}
\label{fig:srvstp2}
\end{figure}

 From the evaluation above, we know that SwitchPairing can achieve secure pairing on two devices. In this Section, we will demonstrate that our SwichPairing also works well on multiple devices. Specifically, three CC2640R2F are involved, and each development board runs our SwitchPairing protocol. All three development boards are connected to the same power source, and we require the volunteer to press switch "on/off" button 4 times.  We set the delay tolerance as 120 ms, 140 ms, 160 ms, 180ms, 200ms respectively for each device.  We run the test for 10 times and evaluate the success rate under different delay tolerance.  Fig.~\ref{fig:srvstp2} shows the results. It can be observed that when the delay tolerance reaches 200ms, the success rate reaches 100\%. Although this delay tolerance is close to (actually below) the median reaction time of a human subject (i.e., 215 ms), it is still challenging for attackers to deploy the ``peep'' attacks, seeing that the success of the attack requires all attacking attempts are made correctly.

Table \ref{tbl:ft} shows the impacts of fault tolerance $\varphi$ on different delay tolerance. It can be observed when the fault tolerance is increased to 25\%, which refers to our protocol allows 1 error to occur (1/4= 25\%), the protocol have a good performance when pair up three devices.  However, we also argue that there is a trade-off when SwitchPairing is used for pairing up multiple devices.
 Manufactures may need to configure the delay tolerance as well as fault tolerance carefully when they want to use our protocol on multiple devices. Moreover, seeing that delay is closely related to chips, manufactures may also enhance the success rate within a small delay tolerance by using a relatively better-designed chip. Another nature solution for this issue is the user may want to make sure that he is not under supervising by an attacker while preforming the pairing on multiple devices, to such a degree that the fault tolerance can be set to a higher value as well, leading to a higher success rate. This is reasonable since standard Bluetooth pairing protocols such as passkey entry make such an assumption.

  \begin{table} 
\centering
\caption{The impact of fault tolerance $\varphi$ and delay tolerance (ms) }
\label{tbl:ft}
\begin{tabular}{ |c|c|c|c| }
\hline
\textbf{\diagbox{ Fault tolerance   }{success rate}{ Delay tolerance   }}  & 120   &  140  &  160  \\ \hline
$\varphi$ = 25\% &    70\%  & 90\% & 100\% \\ \hline
$\varphi$ = 50\% &     100\% & 100\% &  100\% \\
\hline
 \end{tabular}
 \end{table}

 


%% file: Sections/sec8-discussion.tex
\section{Limitations and Comparison}  
\label{sec:disscuss}

We admit that our SwitchPairing protocol also have some limitations:  unlike some context-based pairing protocols such as \cite{miettinen2014context,han2017convoy} that reviewed in Section \ref{subsec:reviewpairing}, which do not interfere users, our pairing protocol requires users' attention while conducting the pairing protocol. 
Regarding this limitation, we argue that (i) our pairing protocol does not require expensive auxiliary equipment/peripherals, which is suitable for the low-priced devices such as smart lamps. That is, our protocol has its advantages when compared to these protocols without consulting a user.   (ii) This drawback widely exists in the most of proximity-based device pairing protocols such as Bluetooth pairing protocol \cite{4.2}. As a type of proximity-based device pairing protocols, our protocol follows the principles strictly, so that our protocol does not downgrade the user-experience of proximity-based device pairing protocols. (iii) The context-based pairing protocols will become defenseless when an attacker is close enough, in that situation the protocol can not distinguish an attacking device from the genuine devices. Our protocol requires the physical access to the victim device, which takes much more efforts for an attacker. Thus, our protocol is more secure than previous protocols.


%% file: Sections/sec9-conclusion.tex
\section{Conclusion}
\label{sec:conclusion}

In this paper, we propose an approach to trade off the security and economics of IoT pairing protocols, which is an issue that most of the existing pairing protocols fail to address. We propose and implement a practical pairing protocol called SwitchPairing, in which two devices pair up together without using auxiliary equipment. The SwitchPairing protocol was demonstrated via two development boards from TI.  
Our implementation shows that our protocol can be extended to other chips and other communicating venues with little effort. Evaluations and user studies are performed to validate this cost-effective solution. Additionally,  our security analysis shows that our protocol is robust against common attacks.

%% file: Sections/sec-acknowledgement.tex
\section*{Acknowledgements}

Jian Weng was partially supported by National Key R\&D Plan of China (Grant Nos. 2017YFB0802203, 2018YFB1003701), National Natural Science Foundation of China (Grant Nos. 61825203, U1736203, 61732021), Guangdong Provincial Special Funds for Applied Technology Research and Development and Transformation of Important Scientific and Technological Achieve (Grant Nos. 2016B010124009 and 2017B010124002). Yue Zhang was partially supported by National Natural Science Foundation of China (Grant Nos. 61877029). Jiasi Weng was partially supported by National Natural Science Foundation of China (Grant Nos. 61802145, 61872153). Zhijian Shao was partially supported by National Natural Science Foundation of China (Grant Nos. 61872153). Ming Li was partially supported by National Natural Science Foundation of China (Grant Nos. 11871248, U1636209). Weiqi Luo was partially supported by National Natural Science Foundation of China (Grant No. 61702222), China Postdoctoral Science Foundation (Grant No. 2017M612842), Postdoctoral Foundation of Jinan University.

%% file: Sections/sec-bio.tex
\begin{IEEEbiography}[{\includegraphics[width=1.1in,height=1.25in,clip,keepaspectratio]{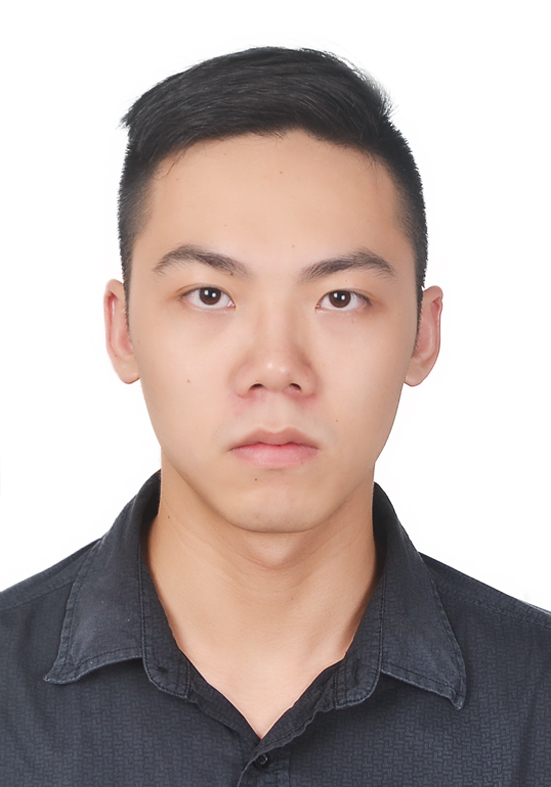}}]{Zhijian Shao}  (Matthew Shao) is a graduate student at Jinan University, China.  His research interest  are IoT security, Virtualization Security and Binary Analysis. Prior to that he received the B.S. Computer Science degree from Jinan University at 2018.
\end{IEEEbiography}

\begin{IEEEbiography}[{\includegraphics[width=1.1in,height=1.25in,clip,keepaspectratio]{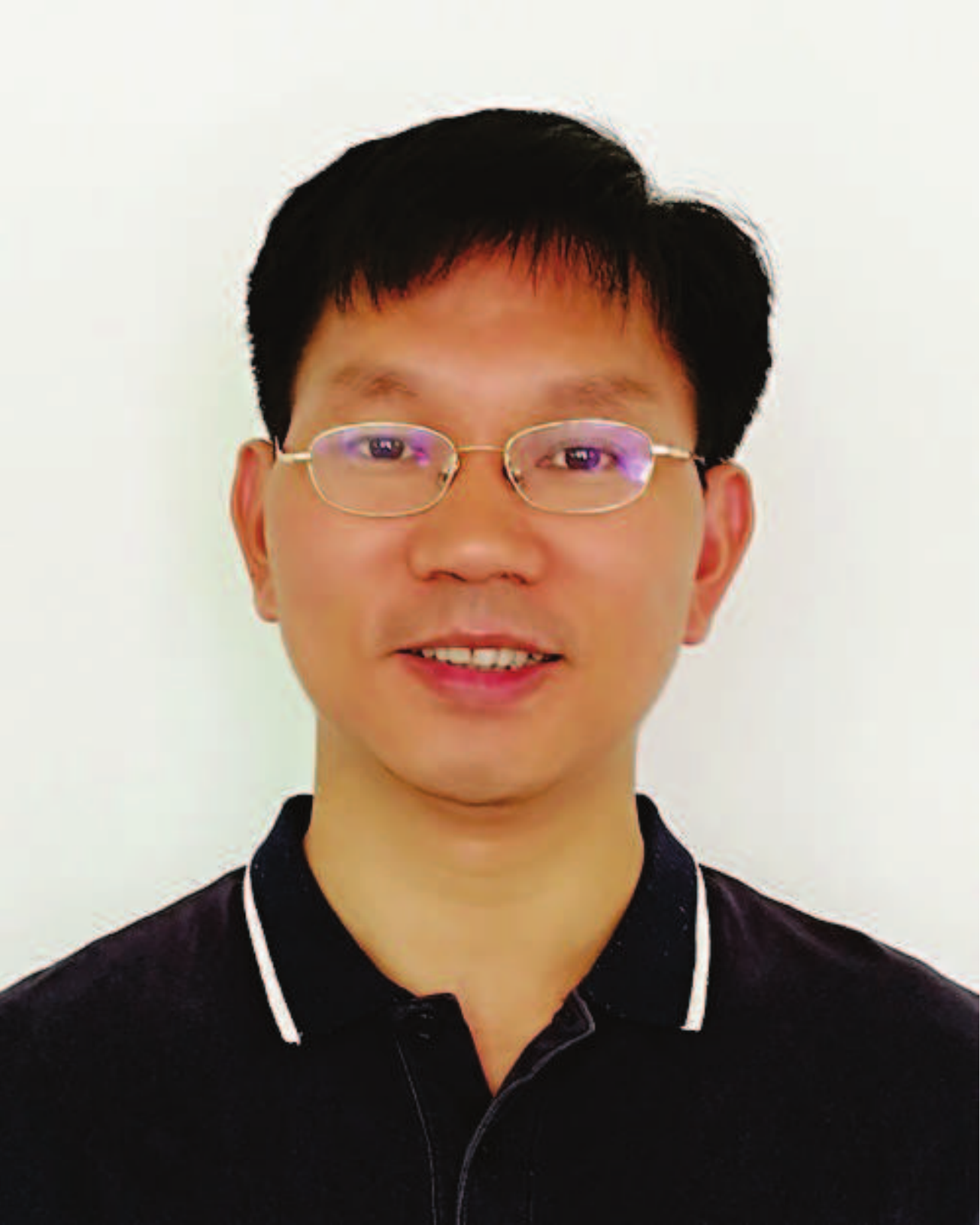}}]{Jian Weng}
is a professor and the Executive Dean with College of Information Science and Technology in Jinan University. He received B.S. degree and M.S. degree at South China University of Technology in 2001 and 2004 respectively, and Ph.D. degree at Shanghai Jiao Tong University in 2008. His research areas include public key cryptography, cloud security, blockchain, etc. He has published 80 papers in international conferences and journals such as CRYPTO, EUROCRYPT, ASIACRYPT, TCC, PKC, CT-RSA, IEEE TDSC, etc. He also serves as associate editor of IEEE Transactions on Vehicular Technology.
\end{IEEEbiography}

\begin{IEEEbiography}[{\includegraphics[width=1in,height=1.25in,clip,keepaspectratio]{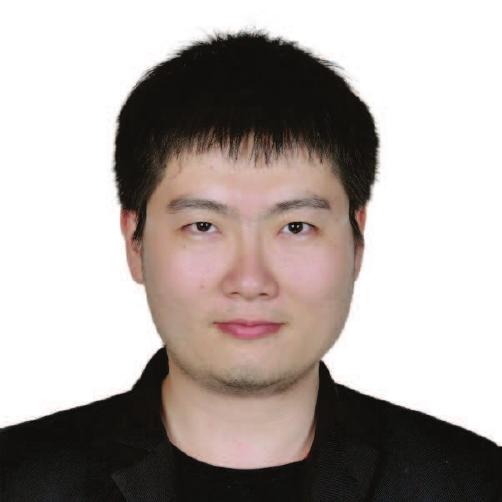}}]{Yue Zhang}
received his B.S. in  information security from  Xi'an University of Posts \& Telecommunications in 2013, and M.S. in information security from  Xi'an University of Posts \& Telecommunications in 2016. From 2016, he started his Ph. D. at Jinan University. His research interests include Bluetooth, system security and Android security.
He has published papers in international conferences and journals such as IEEE TDSC, IEEE TPDS, RAID etc.
\end{IEEEbiography}
 
 \begin{IEEEbiography}[{\includegraphics[width=1in,height=1.25in,clip,keepaspectratio]{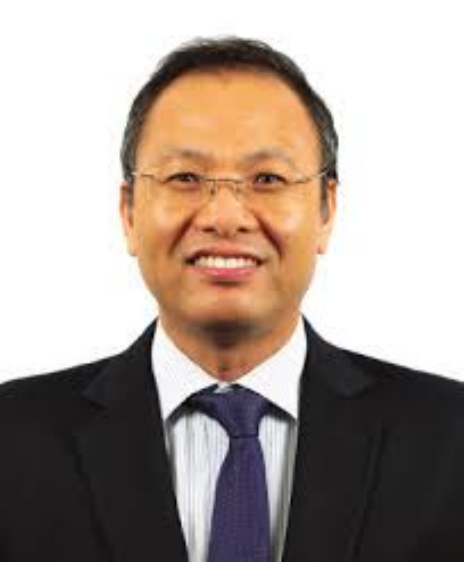}}]{Yongdong WU }
 Received the B.Eng and M.S. from
Beijing University of Aeronautics and Astronautics,
the Ph.D degree from Institute of Automation, Chinese
Academy of Science, and the Master for Management
of Technology from National University of
Singapore. He is a Professor of Jinan University,
China, Adjunct Professor of Wuhan University, China
and CTO of Mirai Electronics Pte Ltd Singapore.
He was Head of System Security Lab, Institute
for Infocomm Research, Singapore. His research
interests include cyber-physical system security, IoT
security, multimedia security, and Network security. He is Associate Editor
of International Journal of Security and Communication Networks, and
was Program co-Chair of the 11th International Conference on Information
Security Practice and Experience (2015). He has published 80 papers as the
first author, and 7 patents, won Tan Kah Kee Young Inventors’ award in 2004
and 2005. His flexible JPEG2000 image protection schemes were incorporated
in the ISO/IEC JPEG 2000 security standard 15444-8 in 2007. He received the
Best Paper Award of IFIP Conference on Communications and Multimedia
Security (CMS) 2012. He was awarded by China-Singapore Joint Research
Programme, NRF (National Research Fund, Singapore), and EMA (Energy
Management Agency, Singapore).
 \end{IEEEbiography}

\begin{IEEEbiography}[{\includegraphics[width=1in,height=1.25in,clip,keepaspectratio]{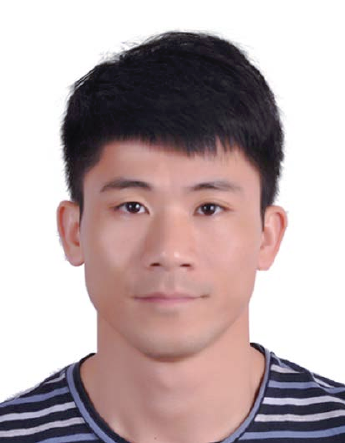}}]{Ming Li}
received his B.S. in electronic information engineering from University of South China in 2009, and M.S. in information processing from Northwestern Polytechnical University in 2012. From 2016, he started his Ph. D. at Jinan University. His research interests include crowdsourcing, blockchain and its privacy and security.
\end{IEEEbiography}

\begin{IEEEbiography}[{\includegraphics[width=1in,height=1.25in,clip,keepaspectratio]{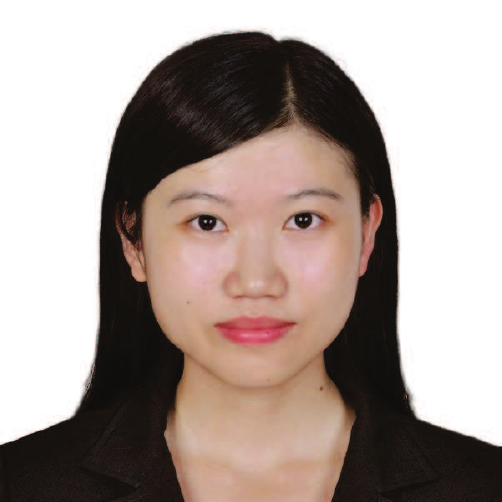}}]{Jiasi Weng} obtained the B.S degree in Software
engineering from South China Agriculture University
in June 2016. She became a graduate student
in Technology of Computer Application from Jinan
University in September 2016. Her research
interests include cryptography and information
security, Blockchain and security in Software Defined Network..
\end{IEEEbiography}
 
  \begin{IEEEbiography}[{\includegraphics[width=1in,height=1.25in,clip,keepaspectratio]{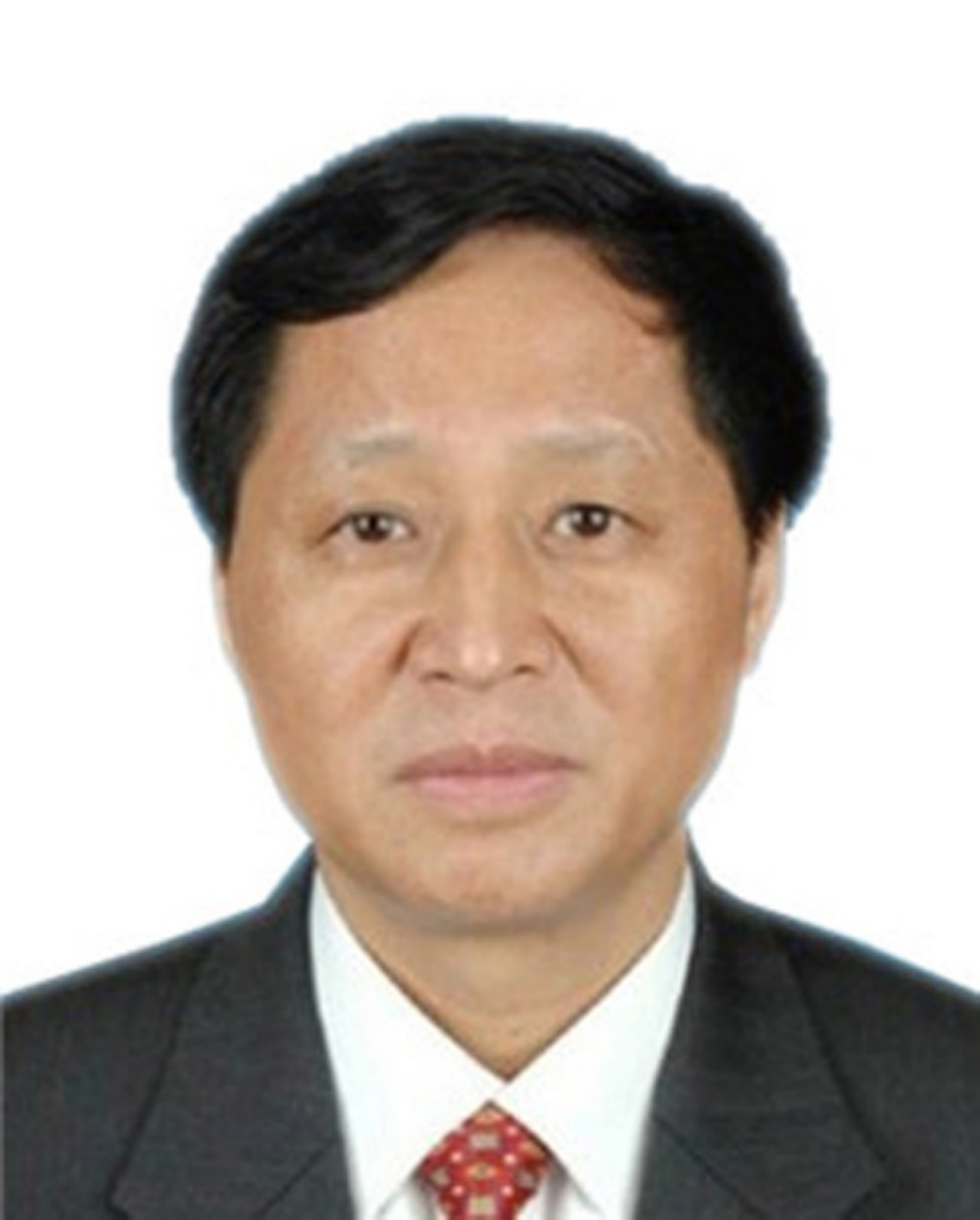}}]{Weiqi Luo}
  received his B.S. degree and M.S.
degree from Jinan University in 1982 and 1985
respectively, and Ph.D. degree from South China
University of Technology in 1999. Currently, he is
a professor with School of Information Science
and Technology in Jinan University. His research
interests include network security, big data, artificial intelligence, etc. He has published more than
100 high-quality papers in international journals
and conferences.
\end{IEEEbiography}

\begin{IEEEbiography}[{\includegraphics[width=1in,height=1.25in,clip,keepaspectratio]{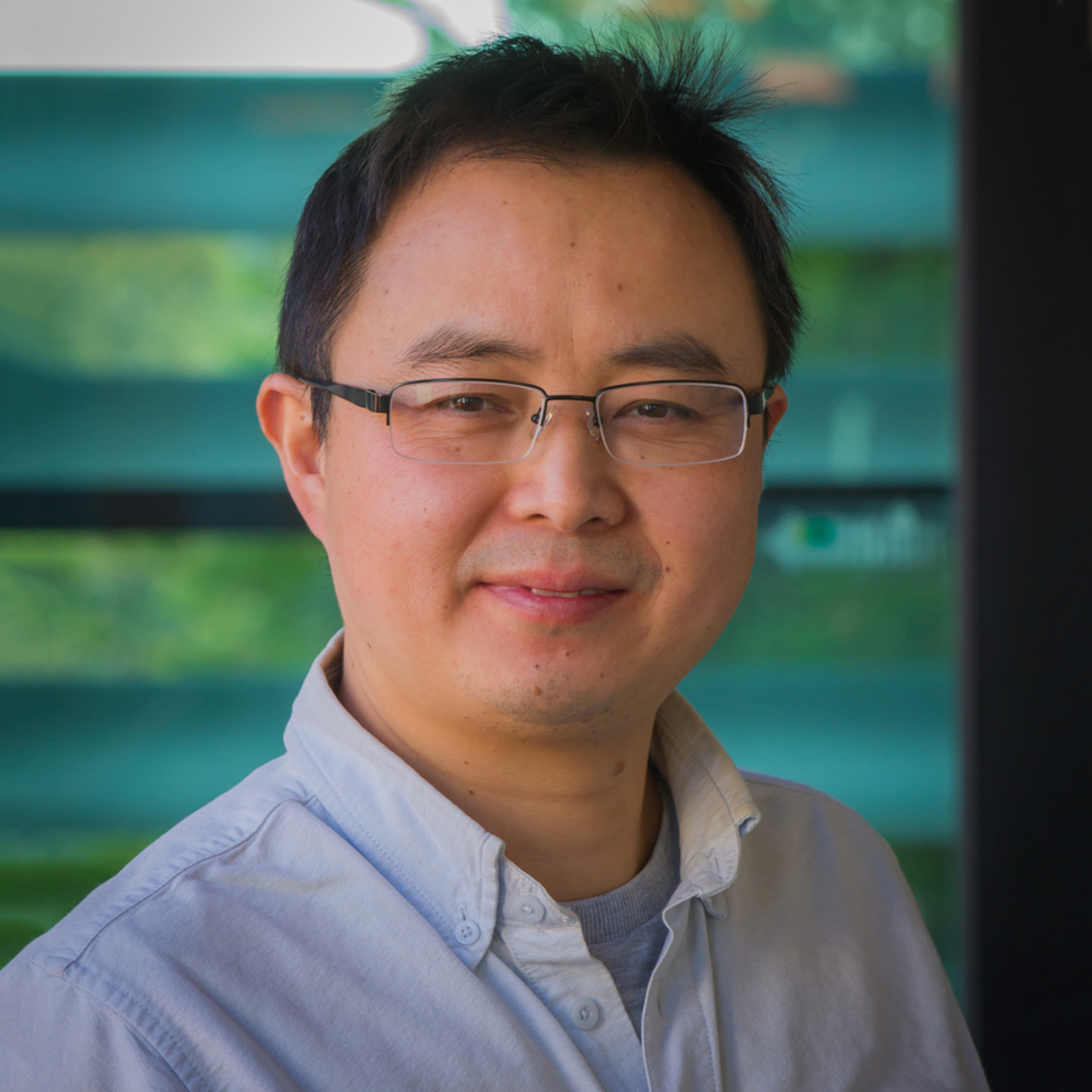}}]{Shui Yu} is currently a Professor of the School of
Software, University of Technology Sydney, Ultimo,
NSW, Australia. He has authored two monographs
and edited two books, over 200 technical papers,
including top journals and top conferences, such
as IEEE TPDS, TC, TIFS, TMC, TKDE, TETC,
ToN, and INFOCOM. He initiated the research field
of networking for big data in 2013. His H-index
is 35. His current research interests include security
and privacy, networking, big data, and mathematical
modeling.
Dr Yu is a member of AAAS and ACM. He actively serves his research
communities in various roles. He is currently serving a number of prestigious editorial boards, including IEEE COMMUNICATIONS SURVEYS AND
TUTORIALS (Area Editor) and IEEE COMMUNICATIONS MAGAZINE (Series
Editor). He has served many international conferences as a member of
organizing committee, such as the Publication Chair for the IEEE Globecom
2015 and the IEEE INFOCOM 2016 and 2017, and the General Chair for
ACSW 2017. He is a Distinguished Lecturer of the IEEE Communication
Society.
\end{IEEEbiography}